\documentclass[bibyear]{aa}    
\usepackage{graphicx}
\usepackage{txfonts}
\usepackage{placeins}                                         
\usepackage{tabularx}    
\usepackage{natbib}

\bibpunct{(}{)}{;}{a}{}{,}

\begin{document}

   \title{AMKID: A large KID-based camera at the APEX telescope}
   \author{N. Reyes\inst{1},
           A. Weiss\inst{1},
           S.J.C. Yates\inst{2}, 
           A.M. Baryshev\inst{1,3},
           I. Cámara-Mayorga\inst{1},
           S. Dabironezare\inst{5,6},
           A. Endo\inst{6},
           L. Ferrari\inst{2}, 
           A. G{\"o}rlitz\inst{1},
           G. Grutzeck\inst{1}, 
           R. G{\"u}sten\inst{1},
           C. Heiter\inst{1},
           S. Heyminck\inst{1},
           S. Hochg{\"u}rtel\inst{1},
           H. Hoevers\inst{5},
           S. Jorquera\inst{4},
           A. Kov\'{a}cs\inst{8}, 
           D. Koopmans\inst{2,3},
           C. K{\"o}nig\inst{1},
           N. Llombart\inst{6},
           K.M. Menten\inst{1},
           V. Murugesan\inst{5},
           M. Ridder\inst{5}, 
           A. Schmitz\inst{1},            
           D.J. Thoen\inst{5},
           A.J. van der Linden\inst{5},
           L. Wang\inst{2,3},
           O. Yurduseven\inst{6},               
           J.J.A. Baselmans\inst{5,6,7}, \and
           B. Klein\inst{1}
           }

   \institute{1 Max Planck Institute f{\"u}r Radioastronomie, Auf dem H{\"u}gel 69, 53121, Bonn, Germany\\
              2 SRON Netherlands Institute for Space Research, Landleven 12, 9747 AD Groningen, Netherlands\\
              3 Kapteyn Astronomical Institute, University of Groningen, The Netherlands \\
              4 Pontificia Universidad Catolica de Chile, San Joaquin, Chile\\
              5 SRON Netherlands Institute for Space Research, Niels Bohrweg 4, 2333 CA Leiden, The Netherlands\\
              6 Faculty of Electrical Engineering, Mathematics and Computer Science, Delft University of Technology, Mekelweg 4, 2628 CD Delft, The Netherlands\\
              7 I.Physikalisches Institut, Universität zu Köln, Zülpicher Straße 77, 50937 Cologne, Germany\\
              8 Center for Astrophysics $|$ Harvard \& Smithsonian, 60 Garden St, Cambridge, MA 02138, U.S.A. \\
              \email{nireyes@mpifr-bonn.mpg.de}\\                                                     
              }
  \date{Received February 10, 2026}
  \abstract
   {Thermal emission at submillimeter wavelengths carries unique information for many astronomical applications, ranging from disks and planet formation around young stars to galaxy evolution studies at cosmological distances. Advancing the mapping speed to detect this faint emission in ground-based astronomy has been a technical challenge for decades.\\
The APEX Microwave Kinetic Inductance Detector (AMKID) camera was designed to accomplish this task. The instrument is a wide field-of-view camera based on kinetic inductance detectors. It is installed on the 12-meter APEX telescope in Chile at 5,100~meters above sea level. The instrument operates in dual color, covering the 350~GHz and 850~GHz atmospheric windows simultaneously. It has a large field of view of 15.3' x 15.3' and an unprecedented number of pixels: 13,952~detectors in the high-frequency band and 3,520~detectors in the low-frequency band. Here we present a complete description of the instrument design and construction, together with results from the successful low-frequency-array (LFA) commissioning campaign executed over the last year.
The LFA performance is in good agreement with design parameters, with detector sensitivity of 2.2~mK$\sqrt{s}$ and diffraction-limited  beam sizes of 17.0''. On-sky measurements demonstrate a sensitivity of 70-90~mJy$\sqrt{s}$ per detector under good atmospheric conditions (PWV below 1.0mm). With this performance the LFA regularly achieves a mapping sensitivity of 25~mJy when mapping a square degree in one hour.   
   AMKID on APEX with its dual-color observing capabilities, high sensitivity, large field of view, and high angular resolution promises to open new scientific opportunities with the APEX telescope.}
         \titlerunning{short title}
   \authorrunning{name(s) of author(s)}
   \maketitle
   \nolinenumbers
\section{Introduction} 
Until its decommissioning in 2020, the wide-field direct-detection camera LABOCA \citep{Kreysa_2003,Siringo2009}, operating up to 295 horn-coupled thermistors in the 870µm atmospheric window, has been the most successful instrument at the APEX telescope \citep{Guesten_2006}, leaving a fantastic legacy of nearly 500 scientific publications. LABOCA's noise-weighted point-source sensitivity per channel of 55 mJy$\sqrt{s}$, after removal of correlated sky noise, served as the gold standard for subsequent developments, but the underlying technology also limited the number of detectors in the camera’s field of view.

A path to more efficient large-scale observations -- toward focal-plane arrays with several kilopixels at affordable costs -- did open in the early 2000s with the introduction of the kinetic inductance detector (KID); \citep{Day_2003}. Since then, a number of KID-based cameras have been deployed for operation in the millimeter regime, such as NIKA and NIKA2 at the IRAM-30m telescope, and MUSIC at the Caltech Submillimeter Observatory \citep{Monfardini_2010,NIKA2,Maloney_2010}. Other instruments covering the submillimeter regime -- such as the TolTec, BLAST-TNG, and P-CAM \citep{Wilson_2020, Lourie_2018, Vaskuri_2025} -- are currently at different stages of development. 

Building on these promises, MPIfR with SRON have launched an ambitious project to develop and operate a wide-field, KID-based submillimeter camera for the APEX telescope: the APEX microwave KID camera (AMKID), initially described in \cite{Esteras_2015}. It is a dual-color instrument, operating simultaneously in the 870 and 350~$\mu$m atmospheric windows, thereby taking advantage of the superb atmospheric conditions at Llano de Chajnantor and the telescope's excellent submillimeter performance. The detectors are lens-coupled MKID devices, arranged in the densest technically possible configuration to fill the telescope's 15.3~arcmin square field of view, with several thousands of pixels detectors per array.

In this article we report on the overall instrument design and focus on the successful commissioning of the low-frequency camera. The high-frequency complement will follow shortly, as will polarization observations.

For a more detailed discussion of the scientific motivations, we refer the reader to the extensive literature  on this subject, which spans disks and planet formation around young stars \citep[e.g.,][]{Andrews_2018,Holland_2017}, large-scale Milky Way studies \citep[e.g.,][]{Schuller_2009,Molinari_2010}, and galaxy evolution at cosmological distances \citep[e.g.,][]{Weiss_2009,Oliver_2012}, among other relevant examples. The promises of mapping large-scale structures on southern skies, be it galactic or cosmological, with the powerful state-of-the-art AMKID arrays installed at the APEX on one of the world’s outstanding submillimeter sites are unique and exciting.\\
\begin{figure}[!h]
   \centering
   \includegraphics[width=\hsize]{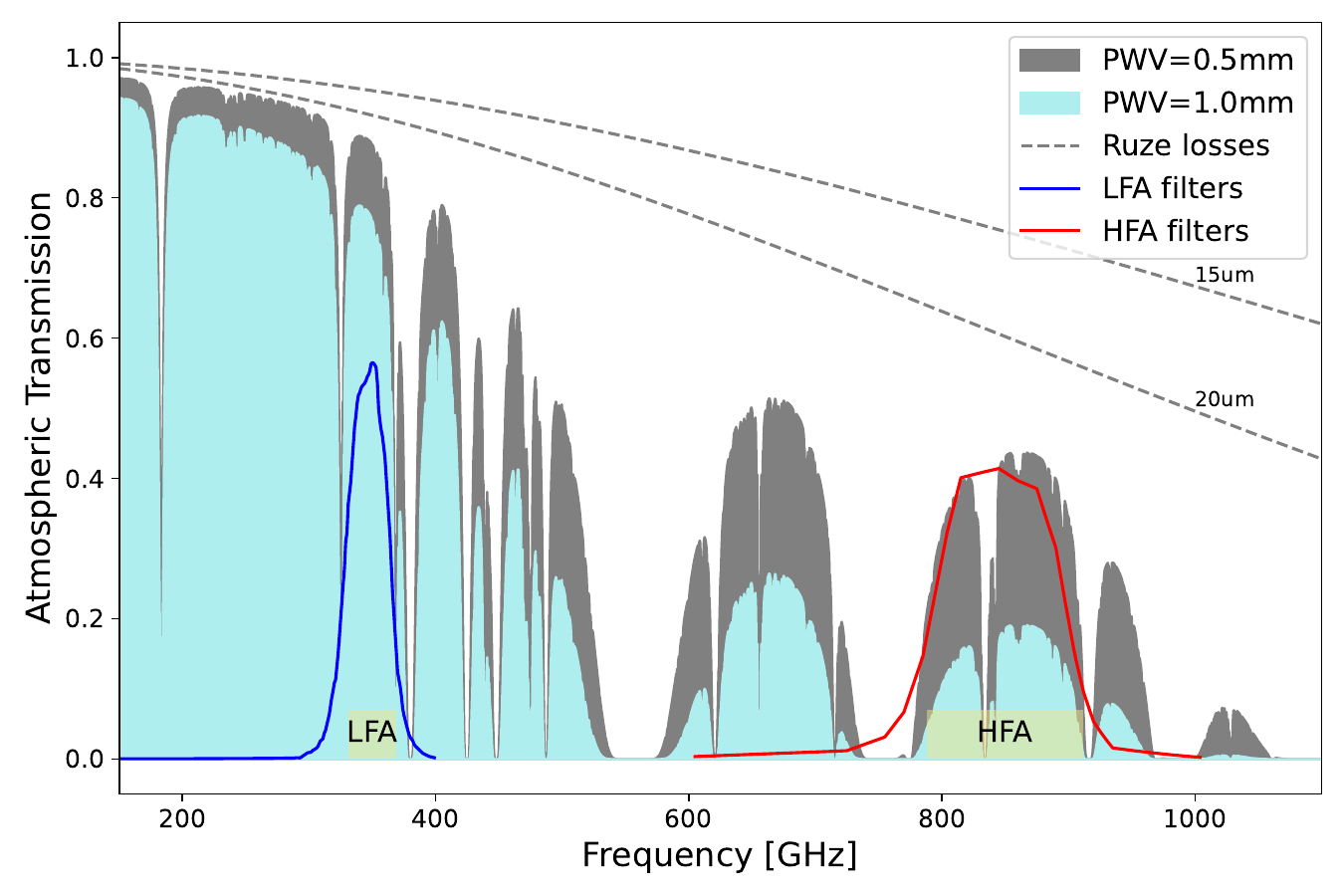}
      \caption{Atmospheric transmission at the Chajnantor plateau. Normal conditions of 1.0mm PWV are regularly observed during several months of the year. Outstanding conditions, below 0.5 mm PWV, are available on average during 890 hours per year. In the same plot, we show the AMKID filter-stack transmission for the two arrays and the Ruze efficiency of a telescope with 15-20 $\mu$m RMS surface quality, a typical value measured via holography at APEX. }
         \label{fig1}
\end{figure}
\section{Instrument overview}
The AMKID instrument is a wide-field submillimeter camera. It operates at 870~$\mu$m (345~GHz) and 350~$\mu$m (850~GHz) using its Low Frequency Array (LFA) and High Frequency Array (HFA) respectively. Both arrays are diplexed in polarization by a free-standing wire polarizer and can be used simultaneously. The instrument is designed for operations at the 12-meter APEX telescope, making use of the superb observation conditions at the Chajnantor plateau. Precipitable water vapor (PWV) below 0.5~mm is observed over several weeks per year, allowing regular operation in the 350~$\mu$m band. In addition, the telescope's surface quality is regularly  monitored and adjusted via holography, maintaining 15-20 $\mu$m RMS over the entire observation season. 

The instrument optics is designed to make use of the maximum square field of view available at the telescope (15.3’~x~15.3’). The field is sampled with an unprecedented number of pixels: 3,520 pixels in the low-frequency band and 13,952 pixels in the high-frequency band, using a pixel spacing of 1.2$\lambda$/D = 18'' at the LFA and 1.5$\lambda$/D = 9'' at the HFA. The instrument optics is near the diffraction limit, with nominal half maximum beam sizes of 17.0’’ at the LFA and 7.5’’ at the HFA.\\

\section{Detector technology} \label{KID_Technology}
   \begin{figure}[t]
   \centering
   \includegraphics[width=0.9\hsize]{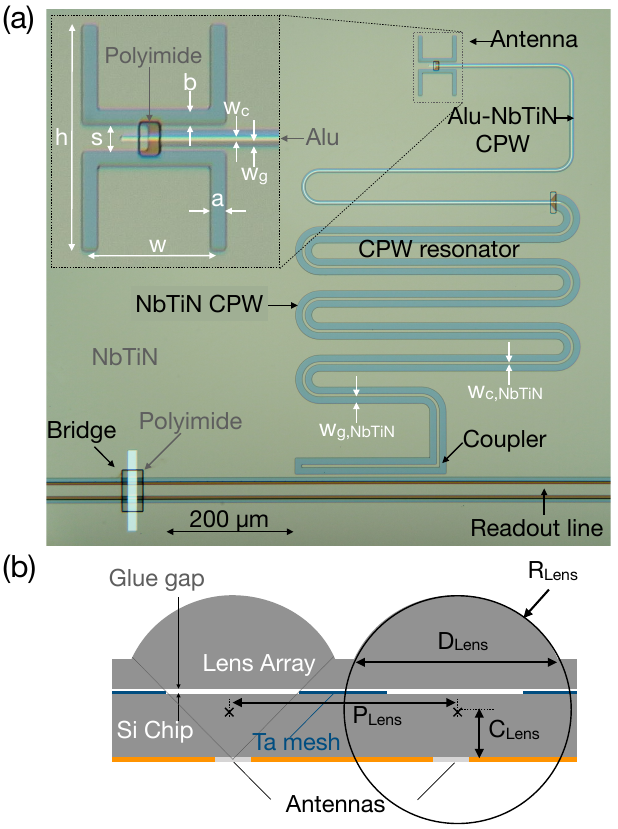}
      \caption{(a) Optical micrograph of a single H band detector with a zoom-in on the antenna structure. (b) Cross section of the chip-lens array assembly. All indicated parameters in panels (a) and (b) are given for both the H band and L band in Table \ref{TableDesigns}.}
         \label{FigKID}
   \end{figure}
        \
        \ \begin{figure*}[t]
   \centering
   \includegraphics[width=1\textwidth]{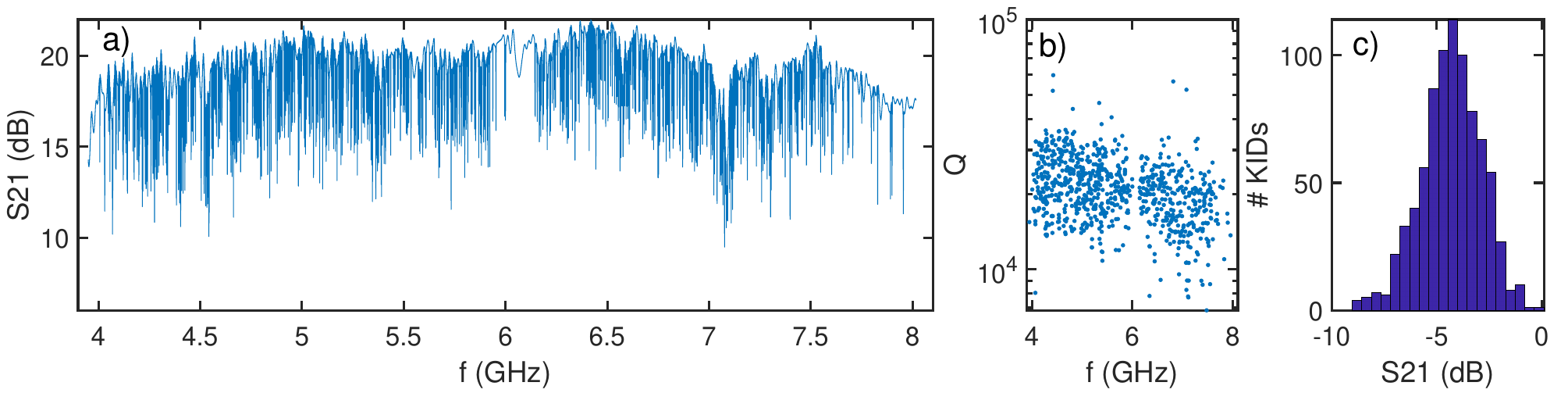}
      \caption{(a) Laboratory frequency sweep of the LT092/LFA4 detector chip, measured against a 300 K load. (b) Measured Q factors; note the slight downward slope with frequency.  (c) Histogram of  KID resonance transmission at center frequency. }
         \label{FigFsweep}
   \end{figure*}        
KIDs were pioneered in the early 2000s by the group of J. Zmuidzinas at Caltech \citep[see][]{Day_2003}. They are superconducting microwave resonators designed to efficiently absorb radiation exceeding their gap frequency (90~GHz for aluminum). Upon radiation absorption Cooper pairs are broken into single particle excitations called quasiparticles, resulting in an increase in both the (kinetic) inductance and the extremely small resistance of the superconducting film at microwave frequencies. As a result, the resonance frequency decreases, which can be read out very accurately using a single microwave probe tone at the unperturbed resonant frequency.

To eliminate thermal excitations and excess noise, KIDs must be operated far below the superconductor critical temperature ($T_c = 1.25$~K for aluminum). At these temperatures all charge transport is achieved by Cooper pairs, which have zero resistance but finite (kinetic) inductance associated with their electron mass. 

The main advantage of this technology is that individual KIDs can be designed with specific resonance frequencies, so that each detector in a KID array corresponds to a single resonance dip, as shown in Fig. \ref{FigFsweep}. As shown, many KIDs can be multiplexed in frequency space and read using a single microwave line, a single cryogenic amplifier, and a single room-temperature backend. This concept allows us to build large arrays, contrary to previous technologies where each detector
required individual wiring that limited the number of pixels.\\

\section{KID detectors fabrication}
The AMKID detectors are NbTiN-aluminum lens-antenna-coupled hybrid KIDs, as described in \citet{Jansen_2013} for the LFA and in \citet{Ferrari_2018a} for the HFA. The detector geometry is shown in Fig. \ref{FigKID}, with dimensions specified in Table \ref{TableDesigns}. They are fabricated on a high-resistivity silicon wafer,  as described in \cite{Thoen_2017} and \cite{Bos_2017}. Both high- and low-frequency detectors are very similar, with the main difference being their antenna dimensions, which scale with operating frequency. This section gives some further details about the detectors' design and fabrication.

The $\lambda/4$ coplanar waveguide (CPW) resonators are made of a 400~nm film of NbTiN with a $T_c = 15$~K. At the open end, the resonator is coupled to the readout line; here, the CPW line is widened to reduce 1/f noise \citep[see][]{Gao2008}. At the shorted end of the resonator, a section of the CPW line narrows significantly, with the central line made of aluminum. The length of this section is 30\% of the KID's length (i.e., about 1-2 mm long, depending on the KID resonant frequency). This scaling improves the readout frequency scatter of the KIDs, since any variations in aluminum line width will have the same fractional effect on all detectors. Another advantage is the frequency-independent responsivity and photon noise level. 

The resonator is coupled to a twin-slot antenna \citep[see][]{Zmuidzinas_1992,Filipovic1993}, which couples the radiation from the telescope into the NbTiN-aluminum hybrid CPW by means of a silicon lens mounted on the chip's backside (see Fig.\ref{FigKID}b). Due to the antenna coupling scheme, AMKID detectors are single-linear-polarization detectors. Quasiparticle creation only occurs in the aluminum line, since the gap frequency of NbTiN is 1.1~THz and thus provides completely loss-free material for the resonator and antenna.

The monolithic silicon lens arrays are fabricated using laser ablation, with a typical fabrication accuracy of $\pm$ 5 $\mu$m in form factor. Each lens array is coated with a $\lambda/4$ antireflection coating made from Parylene-C, as described in \citet{Ji2000}. Numeric simulations restimate detector efficiency at 84\% and 76\% in the low- and high-frequency bands. Between the lens and the planar antenna, we place a capacitive grid that absorbs unwanted in-chip stray light \citep[see][]{Yates_2017}. 

Each detector array is composed of four 61 x 61~mm$^2$ silicon chips with the associated lens arrays, as shown in Fig. \ref{fig:Cryostat}, \ref{NET_Map} and \ref{OpticalCoupling}. The lens arrays and silicon chips are assembled together using dedicated alignment tooling and markers on the lens array and chips. A permanent bond is achieved using cyanoacrylate glue. The LFA chips have 880 detectors each and are read using a single microwave line. On the other side, the HFA silicon chips have 3488 pixels and are read using five coaxial lines, i.e., around 700 detectors per line. \\

\begin{table}[!h]
\caption{Detector design parameters, as defined in Fig. \ref{FigKID}.}
                \small \center
    \begin{tabularx}{0.38\textwidth}{c|c c c c }
                \hline
                \hline
                \textbf{Antenna} & h & a & b & w\\
    \hline 
    LFA & 240 $\mathrm{\mu m}$ &  12 $\mathrm{\mu m}$ &12 $\mathrm{\mu m}$ &25 $\mathrm{\mu m}$  \\
                HFA & 97.7 $\mathrm{\mu m}$ & 6.0 $\mathrm{\mu m}$  & 2 $\mathrm{\mu m}$ & 55.8 $\mathrm{\mu m}$ \\
                \end{tabularx}
                
                \begin{tabularx}{0.45\textwidth}{c| c c c c }
                \hline\hline
                \textbf{Lens } & $\mathrm{R_{Lens}}$ &  $\mathrm{C_{Lens}}$  & $\mathrm{D_{Lens}}$ & $\mathrm{P_{Lens}}$  \\
    \hline 
    LFA & 0.9901 mm & 0.3861 mm & 1.90  mm & 2.00 mm \\
                HFA & 0.5517 mm & 0.2152 mm & 0.940 mm & 1.00 mm \\     
                \end{tabularx}
                
                \begin{tabularx}{0.45\textwidth}{c| c c c c c}
    \hline\hline
                \textbf{KID } & $\mathrm{W_c}$ & $\mathrm{W_g}$  &   $\mathrm{W_{c,NbTiN}}$ & $\mathrm{W_{g,NbTiN}}$ & $Al_{thickness}$ \\
    \hline 
    LFA &  1.6 $\mathrm{\mu m}$ & 2.2 $\mathrm{\mu m}$ & 14 $\mathrm{\mu m}$ & 24 $\mathrm{\mu m}$ & 40 nm\\
                HFA &  1.6 $\mathrm{\mu m}$ & 2.2 $\mathrm{\mu m}$ &  6 $\mathrm{\mu m}$ & 10 $\mathrm{\mu m}$ & 65 nm\\
                \hline
                \hline
                \end{tabularx}\\
                
                \label{TableDesigns}
                \end{table}

To enable small detector sizes, we read out the KIDs in the 4 to 8~GHz frequency band. Resonances are designed to operate with a quality factor ($Q$) on the order of 2.1E4, as seen in Fig. \ref{FigFsweep}b. AMKID operates under significant detector loading powers, of about 4~pW for LFA and 100~pW for HFA, as discussed in Table \ref{table:FilterStackSim}. As a result, the KID internal quality factor ($Q_{i}$) is strongly suppressed, particularly for high-frequency resonators with smaller aluminum volumes. To optimize the noise performance with respect to the readout (see Appendix \ref{DigitalNoise}), resonators are designed to have a depth, $min(S_{21})$, between -3dB and -6dB. The resonance depth is related to the quality factor via $min(S_{21})= Q_c / (Q_i + Q_c)$, with $Q_c$ being the resonance coupling factor. As $1/Q_{i}$ scales with frequency, we choose $Q_c$ to compensate for this effect, while achieving the target KID depth (Fig. \ref{FigFsweep}c).

To maximize the number of KIDs per readout line, the resonant frequencies are designed such that the KID-to-KID frequency spacing increases linearly with frequency, i.e., $\Delta F/F$ is constant. The frequency encoding on the chip ensures that nearest frequency neighbors are always separated by one detector at a completely different frequency \citep{Yates2014}. This prevents electromagnetic cross-coupling and improves frequency scatter, as spatial variations in material parameters vary slowly over medium to large distances.\\
 
\section{Characterization of the detector arrays} \label{ArrayCharacterization}
Detector chips were routinely characterized in the laboratory, using a dedicated test bed described in \citet{Ferrari_2018b}. Measurements were used to verify that the produced detector chip met specifications before integration into the instrument. Here, we would like to highlight some relevant results.

The achieved frequency error ratio ([$F_{design}-F_{measured}]/F_{design}$) for the LFA was on the order of $1\times10^{-2}$, using contact lithography. In contrast, the HFA chips, fabricated later in the project using electron beam lithography for aluminum patterning, exhibit errors on the order of $1.6\times10^{-3}$. Remarkably, this value approaches the best results obtained with chip post-processing as shown in \citet{Shu2018}. As we closely pack our detectors (880~detectors in a 4~GHz bandwidth), we achieve typical frequency spacing between resonators of $\Delta F/F=0.7\times10^{-3}$, still below the fabrication accuracy. This causes the grouping of resonances observed in Fig.~\ref{PowerSweep}. Our calibration scheme recovers most of the affected detectors, but this phenomena causes a loss of around 10\% in yield.

An important characteristic of direct detectors is the noise power spectral density (PSD). This gives the detector output noise, which is a combination of a white, photon noise spectrum and a 1/f term due to two-level systems, as discussed in \cite{Gao2008}. Our detectors are characterized by a 1/f knee frequency of 0.5~Hz. 
A lower knee frequency is possible by increasing detector responsivity through reduced film thickness \citep[see][]{Mazin2004}, although this also diminished the internal Q factor. This would result in broader resonance dips and more resonance frequency collisions per readout line. The current configuration was chosen as a good compromise between the two effects.\\
\section{ Optical system} \label{Optic}
\begin{figure}[ht]
   \centering
   \includegraphics[width=\hsize, trim={4cm 1cm 0cm 1.5cm},clip]{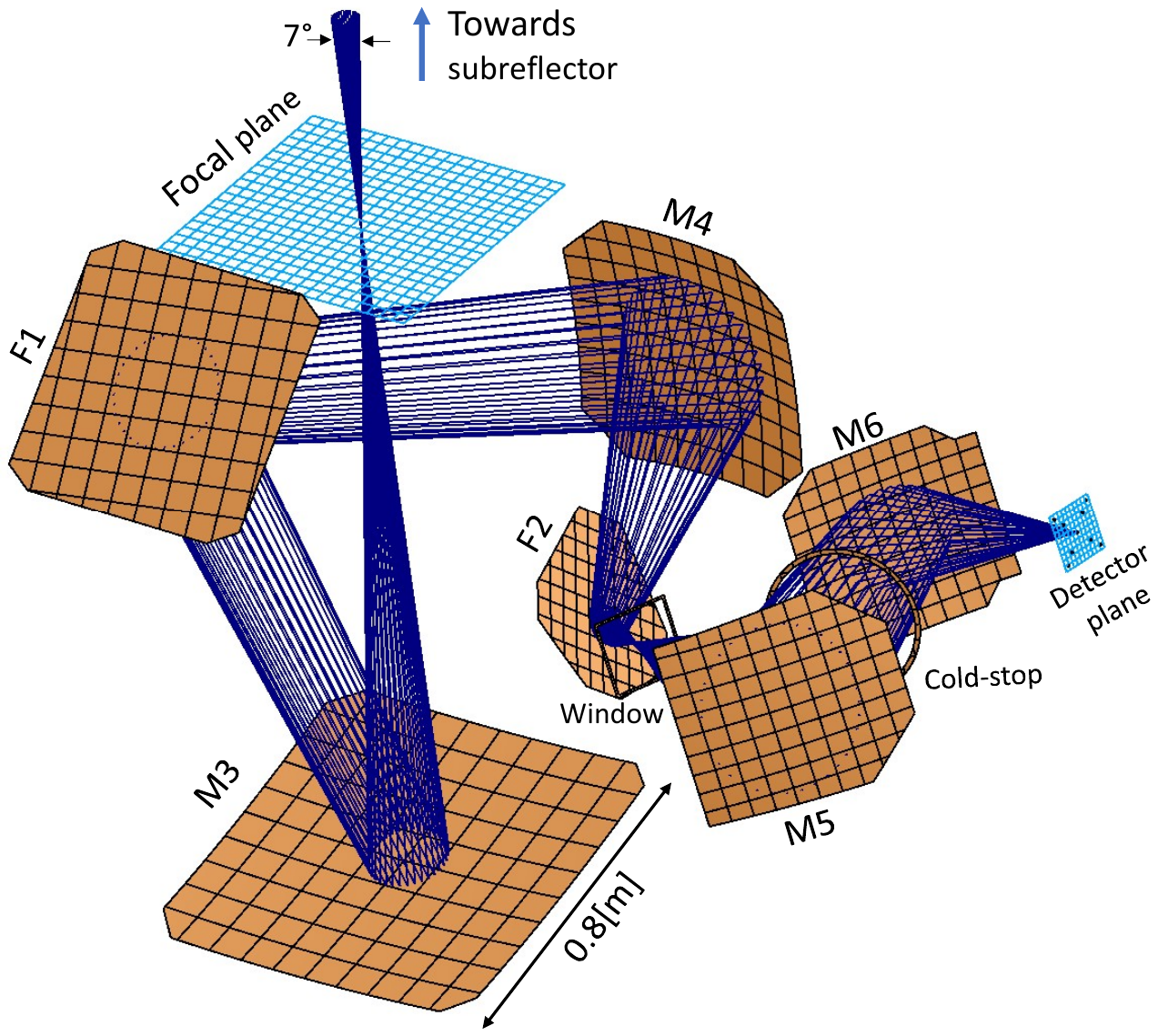}
      \caption{Schematic of the AMKID optics system. It consists of six reflective mirrors, with an overall magnification of 3.6. Only optical rays of the central pixel are shown. }
         \label{fig:OpticSystem}
   \end{figure}

The optical system of AMKID is composed of six reflective mirrors, as depicted in Fig.~\ref{fig:OpticSystem}. The system was designed to use the maximum possible square field of view at the APEX telescope. The initial optical layout, which was later optimized, consists of two Gaussian beam telescopes with a total magnification of 3.6.

The first Gaussian telescope, composed of the mirrors M6 and M5, is located inside the cryostat and has a magnification near unity. The output image plane is used as location for the cryostat window. A cold pupil, at around 8 K,  is located between these two mirrors and shapes the beam to the sub-reflector's size. A cold baffle structure terminates stray light in the system. Electromagnetic simulations (CST studio suite \textsuperscript{\textregistered}) show detector coupling efficiency to the camera pupil of 45\% for LFA and 55\% for HFA. The rather high loss results from the choice of focal plane sampling, which trades off mapping speed versus single-pixel performance through oversampling, as discussed in \cite{Griffin_2002}.

The second Gaussian telescope, consisting of the mirrors M3 and M4, is located outside the cryostat. It converts the image plane at the window location to the telescope focal surface. Telescope boundaries, given by the size and shape of the available space in the Cassegrain cabin, require two additional flat mirrors (F1 and F2) to fold the optic path. Fully reflective optics circumvents the need for critical antireflection lens coating at multiple frequencies. 

This initial design was optimized to compensate for optical aberrations over the entire field of view. Optimization was performed using geometric optics (Zemax OpticStudio\textsuperscript{\textregistered}) by defining all mirror surfaces, including the flat folding mirrors, as extended polynomial expressions. Mirror shape coefficients were optimized to achieve a Strehl ratio greater than 0.8 at 350~$\mu m$ over the entire field of view while ensuring the system is diffraction-limited. A detailed overview of the design process and final results is presented in \cite{Esteras_2015}.\\
Monte Carlo tolerance analysis shows that high-precision mirror alignment relative to the cryostat body is required: $\pm$0.2mm for such critical mirrors as F2 and M4, and $\pm$0.5mm for M3 and F1. Internal cryogenic mirrors are built as a single block and cannot be adjusted. The analysis shows that under these limits the Strehl ratio degradation is lower than 0.95 at 850~GHz. The instrument optics is aligned and regularly verified using a metrology arm (FaroArm\textsuperscript{\textregistered}) installed as part of the instrument in the telescope's Cassegrain cabin.

The optical design was verified using physical optics (GRASP from TICRA) . A set of nine beams across the array was investigated: the four field-of-view corners, the four sub-array centers, and the center position or the system optical axis. For each beam the current distribution over the mirrors was calculated. The cold-stop and window aperture truncation were computed as virtual currents on these apertures. The LFA numeric simulation includes the telescope sub-reflector and the main 12-meter dish. Given numeric complexity, the HFA simulation does not include the antenna. Instead, a  geometric optical model is used to propagate sub-reflector illumination to the main aperture of the telescope. Subsequently, Fourier optics is used to transform the E-field distribution on the telescope aperture into the beam shape on sky. Results confirm that the optical system is diffraction-limited with a beam size of less than 17’’ for the LFA and less than 8’’ for the HFA. Beam ellipticity in both arrays is lower than 1.1 across the entire field of view. 

\section{ Cryogenic system} 
 \begin{figure*}[ht]
   \centering
   \includegraphics[width=1\textwidth, trim={0cm 5cm 0cm 0cm},clip] {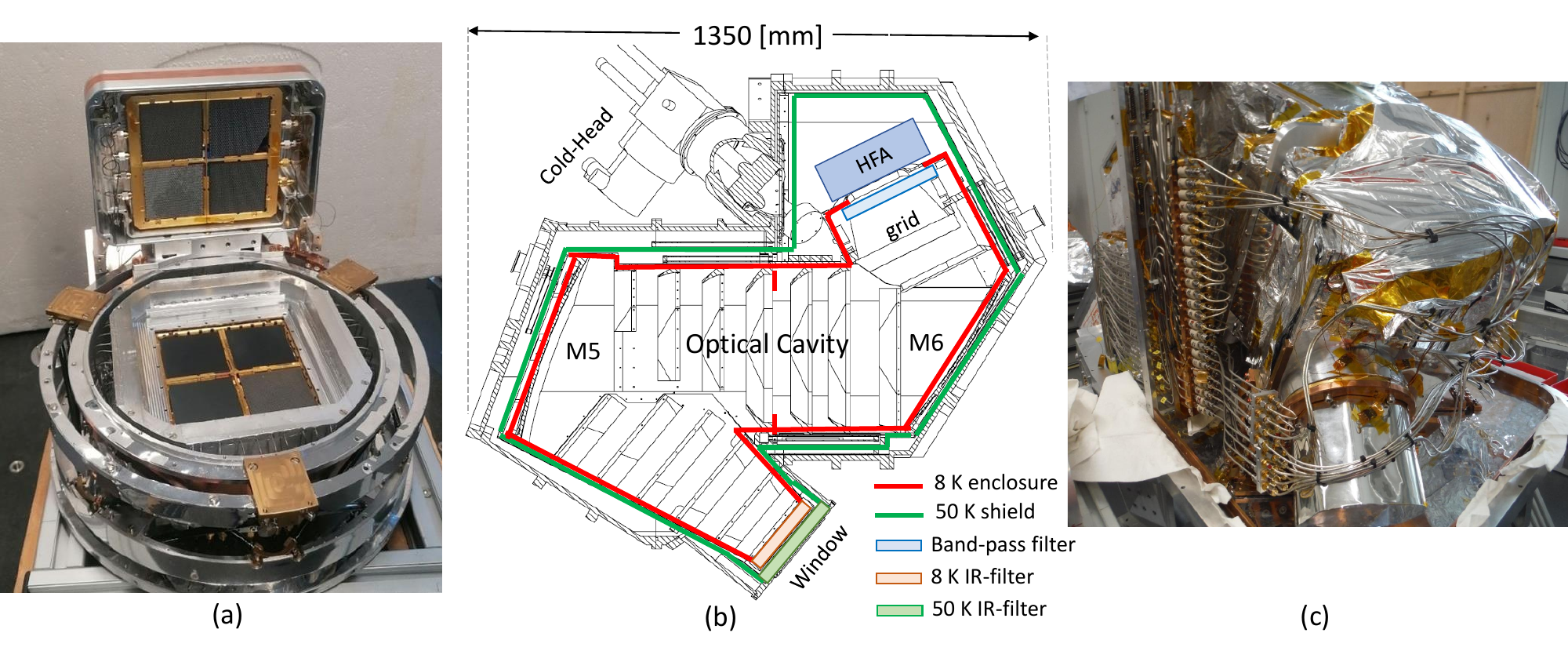}
      \caption{(a) KID detector assembly during installation. The filter stack is removed from the assembly. Hence, the two detectors are clearly visible in the image. The four temperature stages are compactly assembled via a hexapod structure. (b) Cutaway view of the AMKID cryostat. HFA polarization is perpendicular to the image plane. The LFA array is located above the sagittal section. (c) Photography of the assembly and cabling process.}
         \label{fig:Cryostat}
   \end{figure*}        
The KIDs require operation at a temperature low enough that quasiparticle density is dominated by the expected sky-loading power. For AMKID, given the aluminum critical temperature of 1.2~K, this leads to a temperature requirement of 260~mK for the LFA and 280~mK for the HFA. This is achieved using a commercial three-stage sorption cooler. The first stage of the device is based on a $^4$He sorption pump and reaches 1.1 K. The subsequent two stages, both based on $^3$He sorption devices, reach 300 mK and 260 mK under operating conditions. The last stage is used to effectively cool the detectors to the required operating temperature. Under operational conditions the system provides more than 10 hours of holding time, and recycling of the refrigerator system takes about 2 hours. This allows a typical operation time of 20 hours per day. 

Figure~\ref{fig:Cryostat} shows a photo of the structure holding the sorption cooler and the detector arrays. The four temperature stages are compactly assembled via a series of hexapod structures and concentric shields. Superconductive niobium-titanium (NbTi) coaxial cables are used to minimize conductive heat losses between stages. The tight structure and the use of absorbers minimize the stray light radiation reaching the detectors.

The base temperature required to operate the sub-K cooler is provided by a dedicated two-stage pulse tube operating at 3.3~K. A second pulse tube is used to cool the cold optics, the inner optical enclosure (including the baffle structure), and the 24 cryogenic low-noise amplifiers used for instrument readout. The final temperature of this stage is 8~K, with a thermal load of around 4~W. The first stages of both pulse tubes are connected to the infrared shield, achieving a typical temperature between 40 and 50~K. Fig. \ref{fig:Cryostat}b presents a diagram of the cryostat with the main parts described here. The instrument is located in the Cassegrain cabin of the telescope; thus, its mechanical orientation changes with telescope elevation. The system is designed to optimally operate at 50~deg elevation, where the sorption cooler and the pulse tubes are near to vertical orientation. However, all cryogenic subsystems can operate within specifications in the elevation range between 30~deg and 80~deg. Of special concern during operation is the recycle process: maximal efficiency, i.e., maximum holding time, is achieved when recycling is performed in the nominal position, at 50~deg elevation. In particular, the nominal park position, at 11~deg elevation, must be avoided during the recycling process.\\
\section{Filter stack}
KIDs are sensitive to any radiation above the superconductor band gap that reaches the inductive transmission line. Thus, a set of filters is required to define the instrument bandpass. The filter set, which also includes the input cryostat window, plays a relevant role in rejecting infrared and optical radiation.

The cryostat window is located at a highly aberrated image of the focal plane and has a size of 15 x 15~cm$^2$. The window is built from a 5~mm high-density polyethylene (HDPE) plate coated on both sides with a 90~$\mu$m Zitex\textsuperscript{\textregistered} layer serving as an antireflection coating for the 850~GHz frequency range. Given the relative low refraction index of both materials, the reflection losses at 350~GHz are only 5$\%$.

The power that enters the cryostat via the window aperture is around 11~W. The spectrum corresponds to a 300~K blackbody, which peaks in the infrared. This radiation is mostly reflected by a 3~THz low-pass infrared filter located at the 50~K shield, allowing operation of the inner 4~K and lower stages. This filter is designed to be unaffected by self-heating effects due to its reflective design and its implementation as a single-layer structure on a thin substrate; see \cite{Tucker2006}. The remaining infrared power is filtered by a second 5~mm HDPE sheet coated with Zitex installed just after the reflective filter. 

A second filtering step, a 1~THz multilayer low-pass filter, is installed at the 8~K optical enclosure. Thermometers monitor the temperatures at the filter edges (130~K for the HDPE and the reflective filter, 17~K for the 1~THz low-pass). This information feeds our filter stack model described in the next subsection. The large difference between the cold-head temperatures and filters near the window is caused by the large distance separating the two, the large infrared load on our cryostat, and the thermal conduction of the radiation shield structure. 

After the window and infrared filtering stage, the radiation enters the optical cavity, at a nominal temperature of 8~K,  where M5, M6, and the cold stop are installed. The cavity contains a cold-baffle structure that terminates warm stray radiation entering the cryostat from the instrument cabin environment. It also provides cold termination for detector spill-over. Subsequently, radiation reaches the detector area where the free-standing wire grid diplexes both arrays in polarization. One linear polarization is reflected toward the LFA, aligned with detector polarization, while the other polarization is transmitted to the HFA. On both arrays a set of three filters is used to define the passband. The first two, attached to the 4K and 1K stages, are low pass filters. The last filter, located at the 300~mK stage, is a bandpass that defines the operation band of the arrays. Figure \ref{fig1} presents the bandpass of the filter stack, including the cryostat window, used in each array. This information is summarized in Table \ref{table:FilterParameters}.\\
\begin{table}[ht]
\caption{AMKID filter stack parameters.} 
\footnotesize  \centering
\begin{tabular}{c | c c c }      
        \hline\hline              
        Parameter & Central frequency & Bandwidth & Transmission   \\
        \hline                      
                LFA & 350 GHz  &  38 GHz & 56.5\% \\
                HFA &  850 GHz & 123 GHz & 41.4\% \\
        \hline\hline
        \end{tabular}
        \label{table:FilterParameters}
\end{table}

As part of the instrument's design, a model of the filter stack was established. For each filter, the transmitted, reflected, and emitted power were iteratively computed, considering equilibrium with the room-temperature environment and cold-stage temperature. A simple model is used to estimate the power sunk into the thermal stages, but this contribution proves almost negligible. The model requires good knowledge of the transmission coefficients of the filter layers. These were obtained from several sources -- \cite{Tucker2006}, \cite{Ade2006}, \cite{Benford2003}, and \cite{Koller2006} -- along with in-house and third party measurements. After convergence is achieved, the model can estimate several system parameters. The most relevant for our discussion are the overall transmission and bandwidth of the filter stack, the total amount of radiation that reaches the detector, the array noise level, and the main contributors to the noise budget. Table \ref{table:FilterStackSim} presents the main results from this simulation.

Results indicate that the total load (two bands) on our cold detectors is around 4.5 $\mu W$ when observing a 300K load. This value is just within the specification of our sorption cooler. When operating under nominal conditions, i.e., looking at the cold sky, the load diminishes to 1.6 $\mu W$. 

To estimate instrument sensitivity, the sky is modeled as a blackbody source characterized by its brightness temperature. The power emitted by the sky propagates through the system, considering the filters' self-emission. The noise power reaching the arrays is used to calculate the expected noise equivalent temperature (NET) of the system: 0.8~mK$\sqrt{s}$ and 1.2~mK$\sqrt{s}$ for the LFA and HFA. The NET is calculated using the formalism described in Appendix \ref{A:NET}, including the photon shot noise, bunching, and quasiparticle generation-recombination noise in the KID aluminum. Detector two-level system noise is excluded. Of special interest is the identification of main noise contributors. For the HFA the sky noise dominates the noise budget, while for the LFA, sky and filter contributions are comparable. Main noise offenders are the input window for the HFA (8.5\%) and the optical cavity and cold stop (22\%) for the LFA.  It is important to note that this estimation only considers the photon noise and KID noise contributions. The measured NET will increase due to the 1/f noise, the readout noise contribution, and other noise sources.\\
\begin{table}[ht]
        \caption{Filter stack simulation: Load and sensitivity for typical observing conditions.}  
\footnotesize  \centering
        \begin{tabular}{l | c c }      
        \hline\hline
                  & LFA & HFA \\
        \hline
                Sky conditions& PWV=1mm & PWV=0.5mm\\    
                 & el=60deg & el=60deg \\
                Sky temperature   & 70K    & 200K      \\
                Power at array \tablefootmark{a} & 0.06 $\mu$W & 1.57 $\mu$W\\  
                Power per detector & 4pW & 102pW\\
        \hline 
                NET & 0.82 mK$\sqrt{s}$ & 1.16 mK$\sqrt{s}$ \\
                Sky contribution          & 50\% & 79\%  \\
                Stack contribution        & 50\% & 21\%  \\
                \hline    \hline             
        \end{tabular}  
        \tablefoot{
        \tablefoottext{a}{Including polarization grid.}
  }

        \label{table:FilterStackSim}
\end{table}
\section{Readout system} \label{Readout}
\begin{figure}[ht]
   \centering
   \includegraphics[width=\hsize, trim={2cm 0 1cm 0},clip]{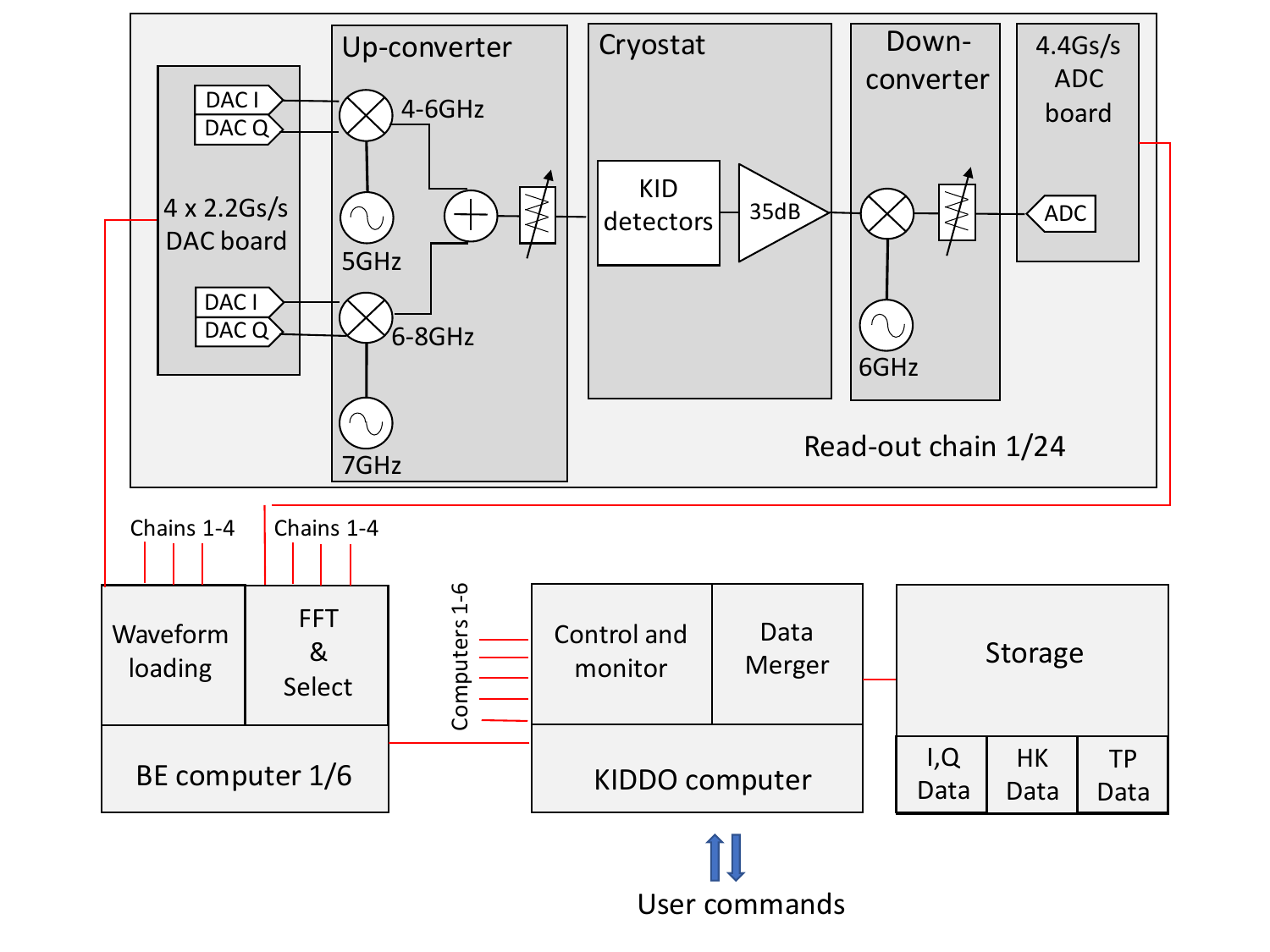}
      \caption{Hardware implementation of the AMKID readout system. }
         \label{fig:Read-out}
   \end{figure}

The readout system, depicted in Fig.~\ref{fig:Read-out}, is designed to measure the frequency shift of the KID resonances under nominal load. For this a comb of frequency tones is used, where each tone is tuned to hit a specific KID. By comparing the input and output signals, the system computes the complex gain coefficient, S$_{21}$, associated with each detector tone. This way, small variations in S$_{21}$ reflect the change in frequency and depth of the associated resonator due to the incoming photons. The S$_{21}$ time streams are stored as raw data in the storage PC using the IQ format: $V_{IQ}=S_{21}\sqrt{50P_{tone}}$, with $P_{tone}$ being the nominal tone power, and 50~$\Omega$ the nominal line impedance. Simultaneously, the system stores the housekeeping data (HK), including tone frequency and power, among other relevant parameters.\\
\indent Figure \ref{fig:Read-out} presents a detailed description of the hardware implementation of the readout system. The architecture uses high-speed digital-to-analog converters (DACs) and digitizer boards (ADCs) to generate and read the required frequency tones for each hardware chain. The procedure can be summarized as follows: Two 14-bit, 2.2~Gsamples/sec DACs in IQ configuration are used to generate an analog waveform with 2~GHz bandwidth, which is up-converted to the 4-6~GHz range. A second set of DACs on the same board generates an additional 2~GHz bandwidth waveform, which is up-converted to the 6-8~GHz range. Both signals are added in a power combiner to produce a 4-8~GHz analog signal containing up to 1080~tones. 

This signal is fed to the KIDs line, and the transmission coefficient associated with each detector is imprinted on the output waveform. The signal is amplified by a cryogenic low-noise amplifier and delivered to a down-conversion stage located outside the cryostat. Here the signal is down-converted to the 0-2~GHz range by a double-sideband mixer. The down-conversion process causes folding of the signal around the local oscillator (LO) frequency, reducing the required sampling rate of the ADC. The main drawback of this technique is the increase in the system noise floor. The possibility of tone collision between the upper and lower sidebands is controlled by the careful placement of the tones. The three LOs involved in the process are tunable, allowing adjustment for differences in the real passband of each detector chain.

The time domain signal is digitized using a 10-bit 4.4~Gsample/sec digitizer and integrated in the FPGA (field programmable gate array) located in the ADC board. Integration is performed using a buffer size of 2$^{16}$ bins, synchronized with the periodicity of the DAC-generated signal. As a result, a perfect sampling of the waveform is obtained. The resulting integrated signal is sent to the back-end computer where a real-to-complex fast Fourier transform (FFT) is applied. Relevant channels of the spectra, where tone information is located, are selected for further analysis and sent to the data merger. The obtained spectral resolution of the system is 67~KHz, enough to place a readout tone close to the KID resonance (typical width of 300-600~KHz) and retrieve the relevant information. The back-end computer also controls waveform generation. Good synchronization of all digital clocks and analog oscillators enables extraction of the amplitude and phase of the measured tones and thus the calculation of the complex gain coefficient for each detector. For this purpose, a high-stability and low-phase-noise signal generator is used to provide a 100~MHz reference signal to all involved hardware devices.

Each back-end computer processes four readout chains, i.e., around 4000~tones. To operate such an instrument, a cluster of six back-end computers is used. All computers send relevant data to the main control computer (KIDDO), where data are merged, formatted, and sent to the storage machine. The maximum speed of the readout system is around 30~samples per second. The main control computer also provides high-level control of the system through configuration files and a user interface based on standard commands for programmable instruments (SCPI).\\
\section{Data processing and calibration} \label{DataProcessing}
The readout outputs the complex gain coefficient (S$_{21}$) for a set of tones placed at the KID resonant frequencies. The S$_{21}$ time streams must be converted into the fractional frequency shift of the KID resonance, $dF/F$, which is proportional to the received power signal for small deviations around the average value \citep[see][]{Calvo2013}. This is the case during regular observations, as the astronomical signal constitutes a small variation above the atmospheric load. This process requires several steps, described in this section.
   \begin{figure}[ht]
   \centering
   \includegraphics[width=\hsize]{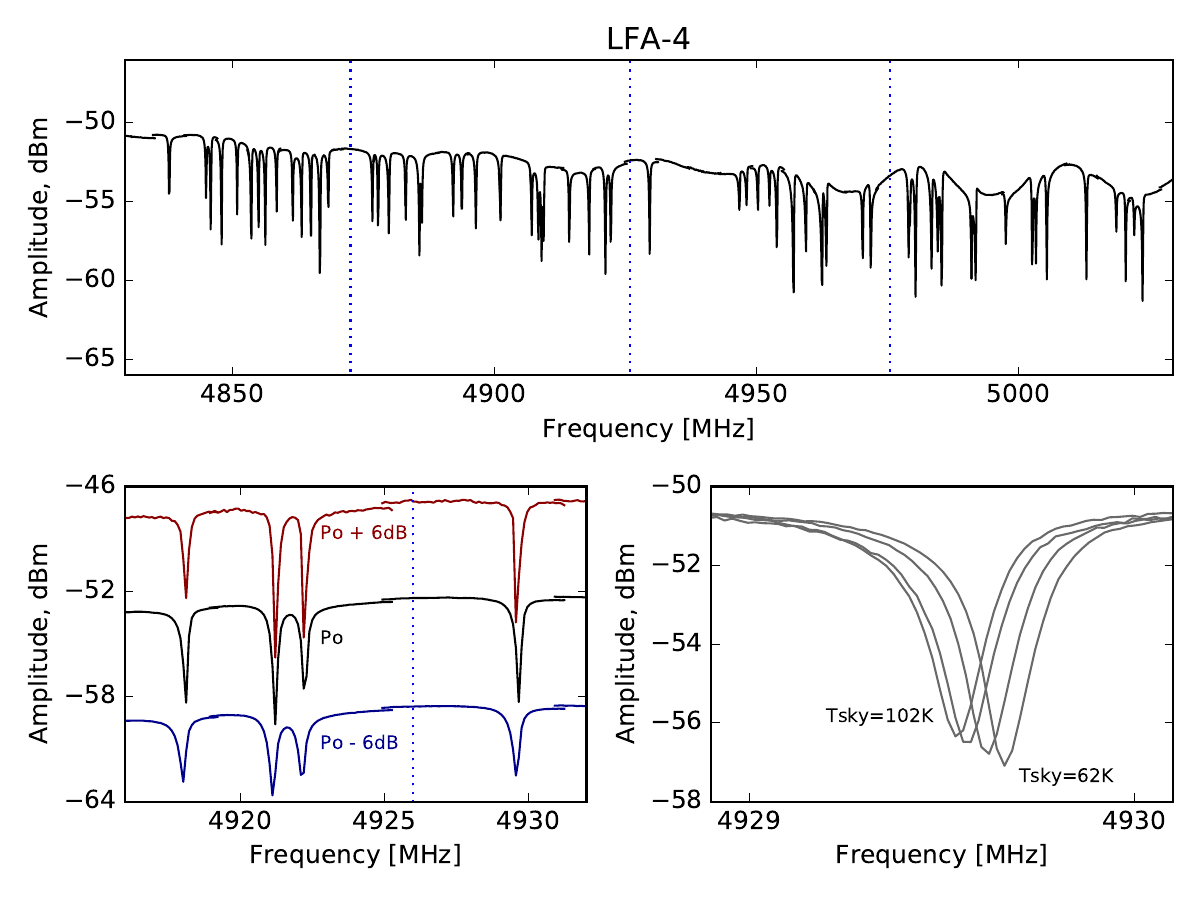}
      \caption{Top: Details of a KID search scan. The data are used to identify resonances and suitable spaces for blind tones (shown by the dashed blue lines). Bottom left: KID search scans repeated at different power levels to find the optimal power for each resonance. Third panel: Response of a KID exposed to different background loads.}
         \label{PowerSweep}
   \end{figure} 
\subsection{Tone list generation and blind tone correction}
The first step consists of finding the resonant frequencies and commanding the backend to place the tones there. Backend functionality includes a search routine that performs a full frequency sweep over the 4-8~GHz range (see Fig.~\ref{PowerSweep} for an example). From this data the appropriate frequency and tone power for each KID are determined. This information is written in the tone list file and loaded into the readout. During this process a set of blind tones, typically 80 per chain, is distributed over the 4-8~GHz bandwidth in frequencies outside the resonances, as described in \cite{Golwala2012}. Signals associated with the blind tones are stable relative to the sky signal and measure time variations in the complex gain of the readout hardware itself. Typical contaminating signals include system drifts, such as as amplifier gain variations and thermal drift in cables.

Considering the up- and down-converter architecture described in Fig.~\ref{fig:Read-out}, blind tones are grouped in quadrants for processing. The data quality of the blind tones is assessed, and noisy blind tones are discarded. Selected blind tones are normalized, phase-referenced and averaged per quadrant. The result is used as a correction factor for the measured S$_{21}$ time-ordered data.

Apart from correcting all system drifts, this blind tone correction also improves the effective noise floor of the readout electronics. In particular, it effectively removes the large phase-noise contribution, of around 30~dB, from the three LOs in the conversion scheme. As a result, the system achieves a typical noise floor value close to -93~dBc/Hz when operating with 1000 tones. As the typical KID signal is in the range of -90~dBc/Hz, we estimate the readout noise contribution to the instrument NET as an additional 15\%, as discussed in Appendix \ref{DigitalNoise}.\\
         \begin{figure}[ht]
   \centering
   \includegraphics[width=\hsize]{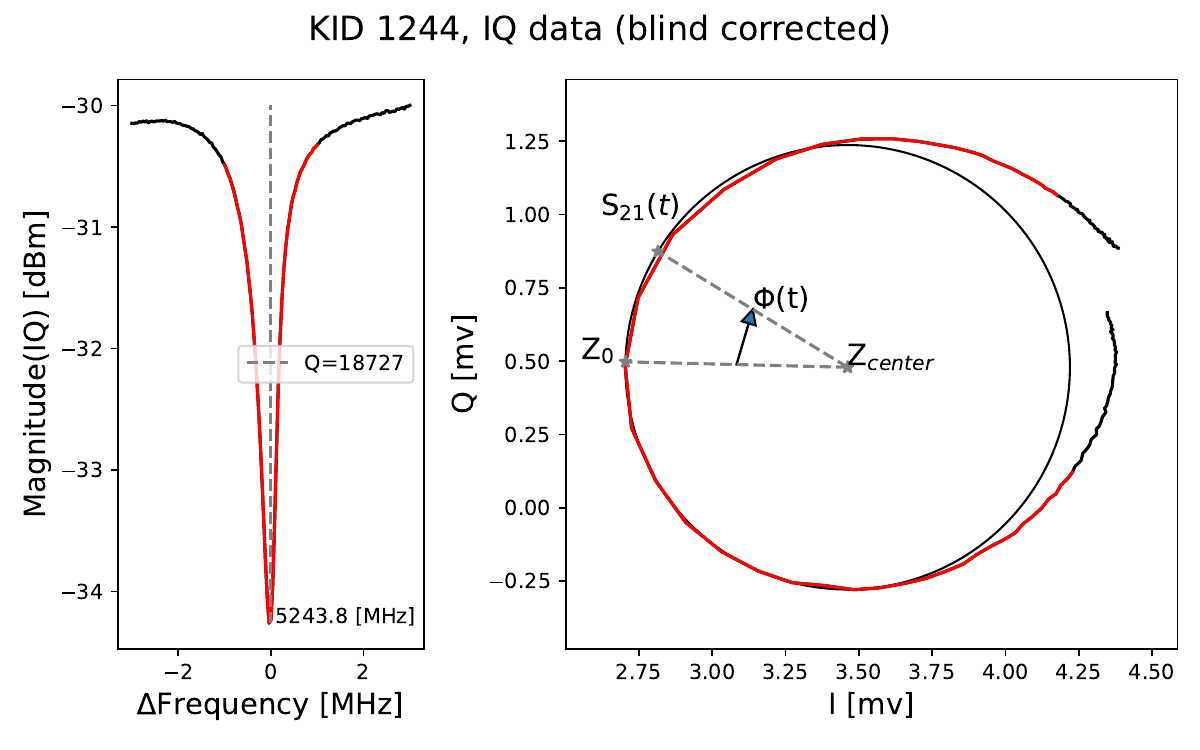}
      \caption{Calibration scan of a single LFA KID. Left: Amplitude of the IQ data where the resonance is clearly observed. Right: IQ data in the complex plane. Here the resonance is represented as a circle. Circle parameters are also depicted in the figure. }
         \label{fig10}
   \end{figure} 
\subsection{Circle calibration, frequency sweep, and linearization}
\begin{figure}[ht]
   \centering
   \includegraphics[width=\hsize]{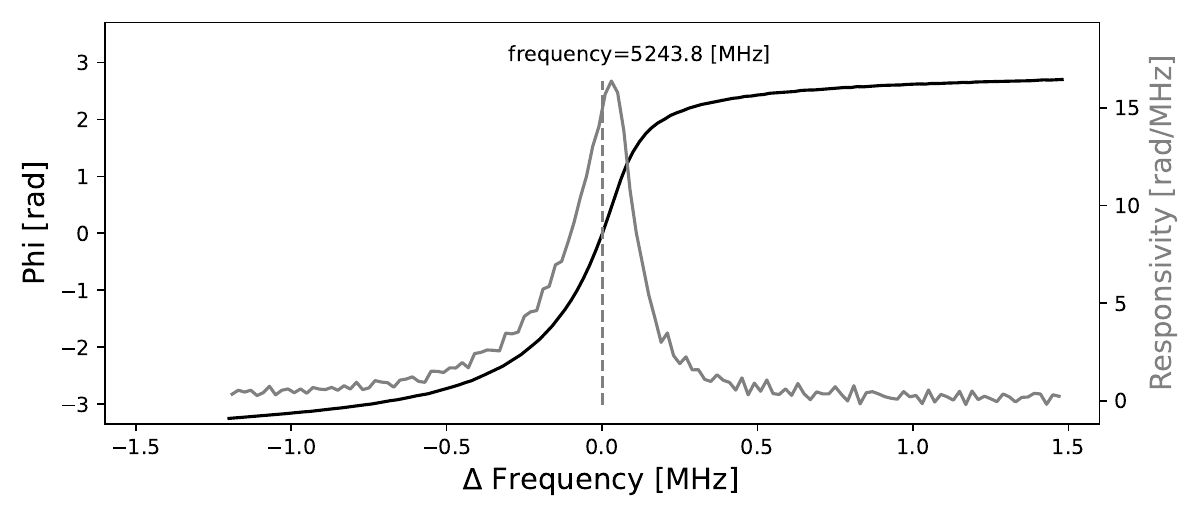}
      \caption{Zoom-in on specific KID data. The solid black trace shows the relation between phase shift ($\Phi$) and frequency shift ($dF$). The gray trace shows the detector responsivity. Optimal performance occurs when the readout tone is placed at maximum responsivity, as indicated in the figure. }
         \label{fig11}
   \end{figure}

After blind-tone correction, the S$_{21}$ timelines are converted into frequency changes of the KID resonances ($dF$). This process requires that KID resonances are well characterized in the complex plane, i.e., amplitude and phase. This characterization is done via a frequency sweep. It consists of placing the readout tones on each resonance, followed by synchronously sweeping the LOs of the up- and down-converters over the range from -1.5 to 1.5~MHz. As a consequence, readout tones sweep around its nominal value within the same frequency range. An example of such a measurement is presented in Fig.~\ref{fig10}. The result is used to create a full characterization of the KID resonances, which is stored by the analysis software and used to calibrate subsequent regular data. As seen in Fig.~\ref{fig10}, resonances in the complex plane correspond to circles and can be characterized  by storing the circle center (Z$_{\rm center}$), circle radius ($R_0$), and the phase-zero point ($Z_0$). This simple three-parameter model is known as the "KID circle." To reduce observational data, the blind-tone-corrected S$_{21}$ values are mapped onto the calibration circle using
\small
\begin{equation}
\Phi(t)= angle\left( \frac{S_{21}(t) - Z_{\rm center}}{Z_0 - Z_{\rm center}} \right) 
\label{eq1},
\end{equation}
\normalsize   
where $\Phi(t)$ corresponds to the phase shift in the KID circle between the phase-zero point and the current measurement. For small signals on a large background, the phase shift is proportional to the fractional frequency shift in the KID resonance, $dF(t)/F$, as shown by  \cite{Calvo2013} and \cite{Bisigello2016}. Therefore, it can be used to measure the detected power.\\
In the general case, the conversion from $\Phi(t)$ to $dF(t)$ is obtained by inverting the $\Phi-dF$ relation obtained during the frequency sweep. Figure~\ref{fig11} demonstrates this relation for a specific resonance. In the plot the linear regime is clearly seen, together with the responsivity of the detectors at different tone placements or biases. 
\subsection{Temperature calibration} \label{WireScanner}
To create an absolute temperature calibration scale, we use a wire scanner. This is a device that moves an orthogonal pair of wires through the instrument focal plane. The first wire moves along the $\hat{x}$ direction, while the second scans along the $\hat{y}$ direction. The warm wire movement against the cold sky produces a small signal on the KID detectors, used to establish the positions of each detector in the focal plane and convert each detector's response from frequency shift to antenna temperature.

The wire scanner was originally developed for laboratory characterization, as described in Appendix \ref{APENDIX1}. However, since it allows us to characterize individual detectors' responses to a known signal in just 90 seconds, it has been implemented in the AMKD calibration theme during regular observations. Besides providing temperature calibration and thus camera flat-fielding, it enables monitoring the instrument, assessing tone frequency adjustment needs, and subsequent circle calibration over time. The coupling of individual detectors to the telescope is measured via beam maps on planets, providing the conversion from the temperature scale to flux density. This quality only depends on the optical alignment of the camera and is stable over time.\\
A wire size of 0.5~mm was chosen to provide a small signal of around 10~K, which ensures that detectors are kept in their linear regime during measurement. Unfortunately, the required wire diameter is similar to the operation wavelength: 350~$\mu$m at the HFA and 870~$\mu$m at the LFA. As a result, diffraction and polarization effects must be understood \citep{Xie_1991}, especially in the low frequency regime where the wavelength exceeds the wire thickness. Polarization effects average out if we take two orthogonal scan directions, as we usually do when retrieving detector positions. For diffraction effects we calibrate the effective wire peak temperature by using a wire grid rotator that creates a deterministic absolute temperature scale using two well known calibration sources (liquid nitrogen and ambient load). This information was used to retrieve an effective wire diameter ($d_{wire}$) that accounts for diffraction effects, assuming the wire is in thermal equilibrium with the ambient temperature ($T_{wire} = T_{ambient}$). As expected, diffraction effects are only relevant for longer wavelengths, as we find $d_{wire}=0.6$~mm for the LFA and $d_{wire}=0.5$~mm for the HFA.  
   \begin{figure}[ht]
   \centering
   \includegraphics[width=\hsize]{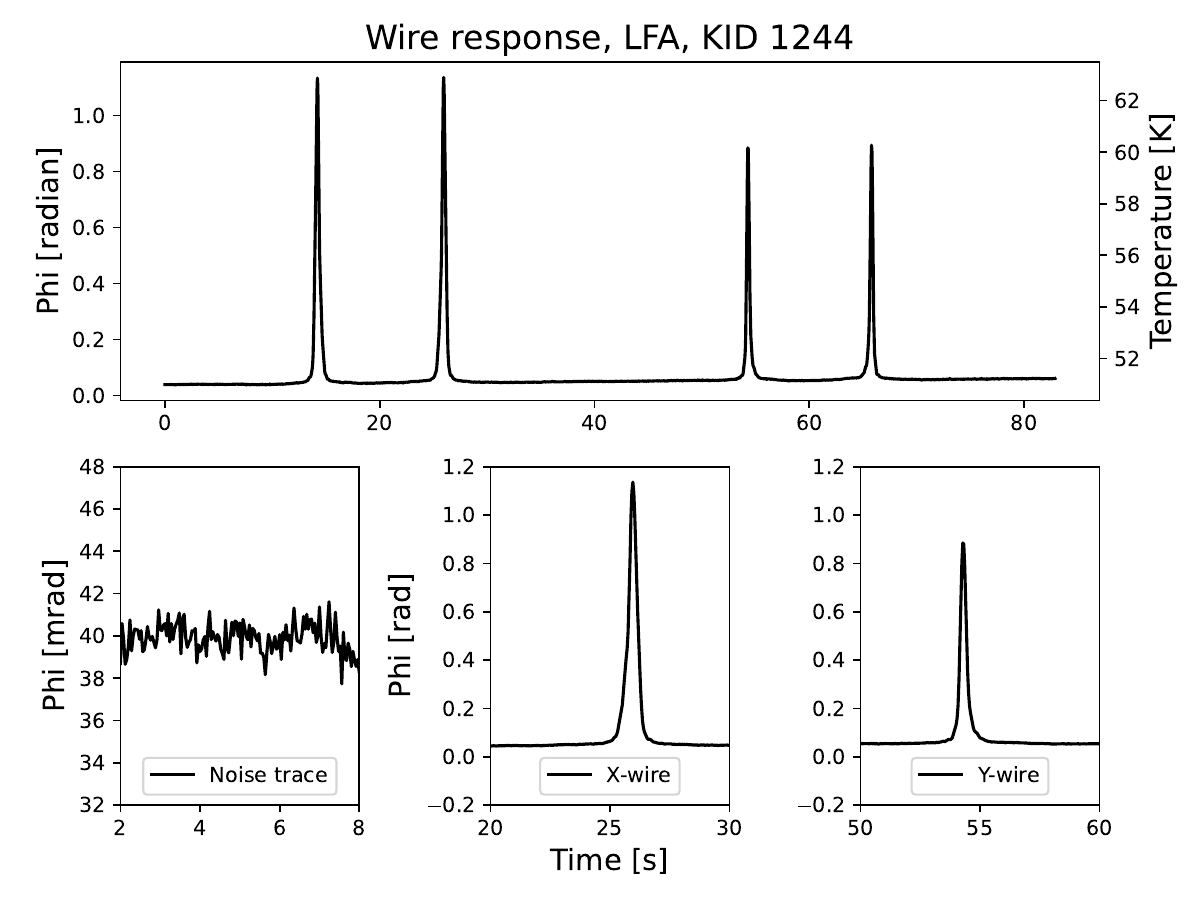}
      \caption{Top: 870$\mu$m detector response to a wire movement in the focal plane. The first two peaks correspond to the inward and outward movement of the wire aligned in the $\hat{x}$ direction. The next two peaks are the response to the $\hat{y}$ direction wire. Bottom: Zoom-in on the quiet part of the trace used for noise statistic calculation, followed by a zoom-in on an $\hat{x}$-wire and $\hat{y}$-wire response. The different amplitudes are due to the different polarization alignments. }
         \label{fig_WireScan}
         \end{figure}
Figure~\ref{fig_WireScan} shows the typical time response of a low-frequency KID detector to the cable movement. An interesting feature of this signal is that the peak response depends on the optics status: maximum response occurs when the wire is moved across the detector's focal point. The key point for calibration is that an aberrated or defocused beam reduces the peak response but widens it. Overall, the integral of the trace remains constant, as long as the detector operates in its linear regime and no extended detector responses exist \citep[as seen in][]{Yates_2017}. 

The wire signal is used to obtain the responsivity of each detector and flat field the camera response. This is implemented in the conversion of data to the temperature scale using
\small
\begin{equation}
K=\frac{dF}{\Delta T}=\frac{\int{Wire_{trace}dx}}{(T_{wire} - T_{sky})d_{wire}}
\label{eq2},
\end{equation}
\normalsize
where the integral in the numerator is the integral strength of the wire response, $T_{wire}$ is the physical temperature of the wire, $T_{sky}$ is the sky brightness temperature obtained from 183~GHz radiometer measurements, and $d_{wire}$ is the effective size of the wire. Once the detector response is calibrated, part of the timeline can be used as a noise trace (figure \ref{fig_WireScan}.b). The noise trace is processed to remove correlated signals, and the NET on blank sky is calculated as\small
\begin{equation}
NET=\frac{\sigma(Noise_{trace})}{\sqrt{f_{sampling}}} \quad [K\sqrt{s}]
\label{eq3},
\end{equation}
\normalsize
where $f_{sampling}$ is the sampling frequency of our readout. 

An interesting application of the wire scanner is the optimization of the readout power for each resonance. In this process several circle calibrations followed by wire scanner measurements are performed at different readout power levels. Generated data enable optimization of individual power levels to maximize detector sensitivities. We observe that adjacent (in frequency space) KIDs can require power differences as large as 8~dB; optimal performance is obtained with a power adjustment resolution of 0.5dB.\\

\section{Instrument performance}
After the extensive laboratory characterization described in Appendix \ref{APENDIX1}, AMKID was shipped to Chile and installed in its current configuration in August 2023. The process included the in situ assembly of the last generation HFA detectors and the room temperature magnetic shield described in Appendix \ref{Ap:Magnetic}. After installation, several calibration sources were observed using the dual-color configuration. These initial observations prove that the observing strategies, pointing model, and beam quality are adequate for both arrays. During the 2025 observing season, work focused on the complete science verification and optimization of the LFA. High-frequency array verification remains ongoing, and final results are expected at the end of the 2026 observing campaign.

In the following sections, we present the main results obtained from the successful AMKID-LFA commissioning. This section focuses on measured instrument performance, while Sect. \ref{On-Sky} presents early on-sky observations that prove basic instrument capabilities. \\
\subsection{Array sensitivity}
\begin{figure}[t]
   \centering
   \includegraphics[width=\hsize]{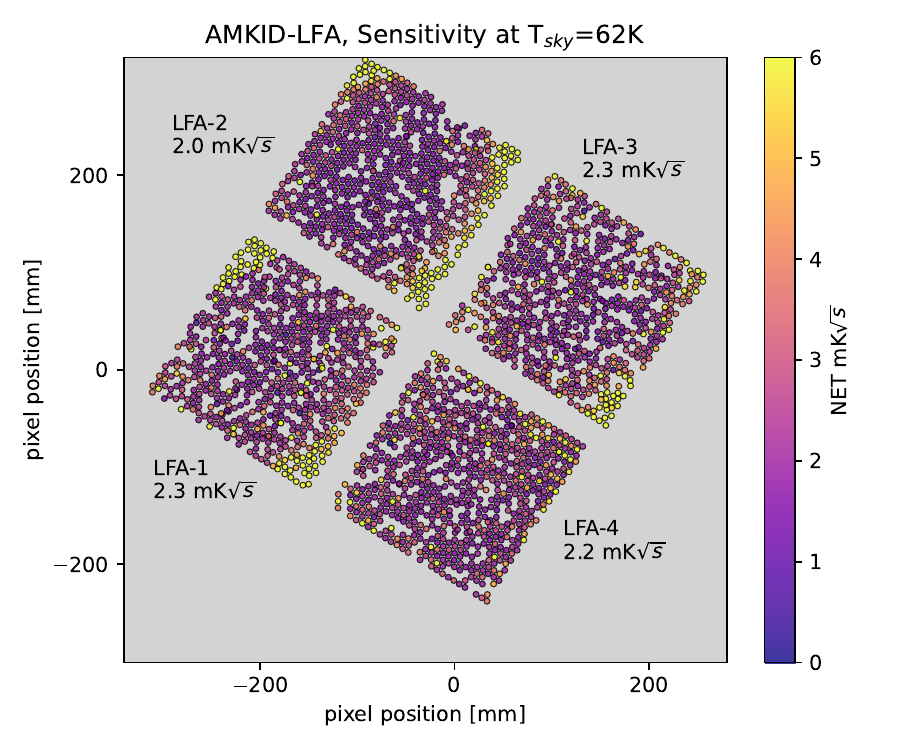}
      \caption{Spatial distribution of NET for the LFA detectors. The median NET of each detector chip is annotated, together with the ID name assigned to each readout line. Defective corners are mostly caused by problems in the lens array gluing process.}
         \label{NET_Map}
   \end{figure}

The array NET is regularly measured using the wire scanner technique. Figure~\ref{NET_Map} shows the results of a typical measurement taken during LFA observations. It can be observed that most of the array exhibits uniform performance around 2.2~$mK\sqrt{s}$ (median value). Some detector chip corners show degraded performance, especially on chip LFA-2. This effect is caused by problems in the lens array gluing process. On LFA-1 a small broken corner in the array appears as an abrupt change in the NET value, contrary to the observed degradation in other corners.\\
\indent For such a high number of pixels, statistical analysis is mandatory; see Fig. \ref{fig:Hystograms}. The distribution mode represents the performance of well-behaved detectors and can be directly compared with our modeled sensitivity. By contrast, the median value includes nonideal detectors in the array and is therefore useful for estimating final mapping speed.

When comparing the measured (mode value is 1.7mK$\sqrt{s}$) with our predicted NET values (0.8mK$\sqrt{s}$), we observe a large discrepancy. This is further discussed in Sect. \ref{Stability}. The main conclusion is that this difference is explained by several factors: 1/f noise contribution, contaminating signals, and digital readout contribution.

The spread in the NET distribution degrades camera performance. In an ideal instrument, the median would approach the mode value, as most detectors would achieve the same performance. The observed spread is caused by several factors: chip performance degradation toward the edges, imperfect resonance biasing, and increased readout noise contribution at high frequencies. A new generation of readout technology, currently under development, will help control the last two factors.

It is important to note that when initially deploying the instrument, sensitivity spread was a major problem. Implementing an accurate method to optimize resonance bias reduced the NET median value by 20\%.\\
\subsection{Stability} \label{Stability}
\begin{figure}[ht]
   \centering
   \includegraphics[width=\hsize, trim={1cm 0 1.5cm 0},clip]{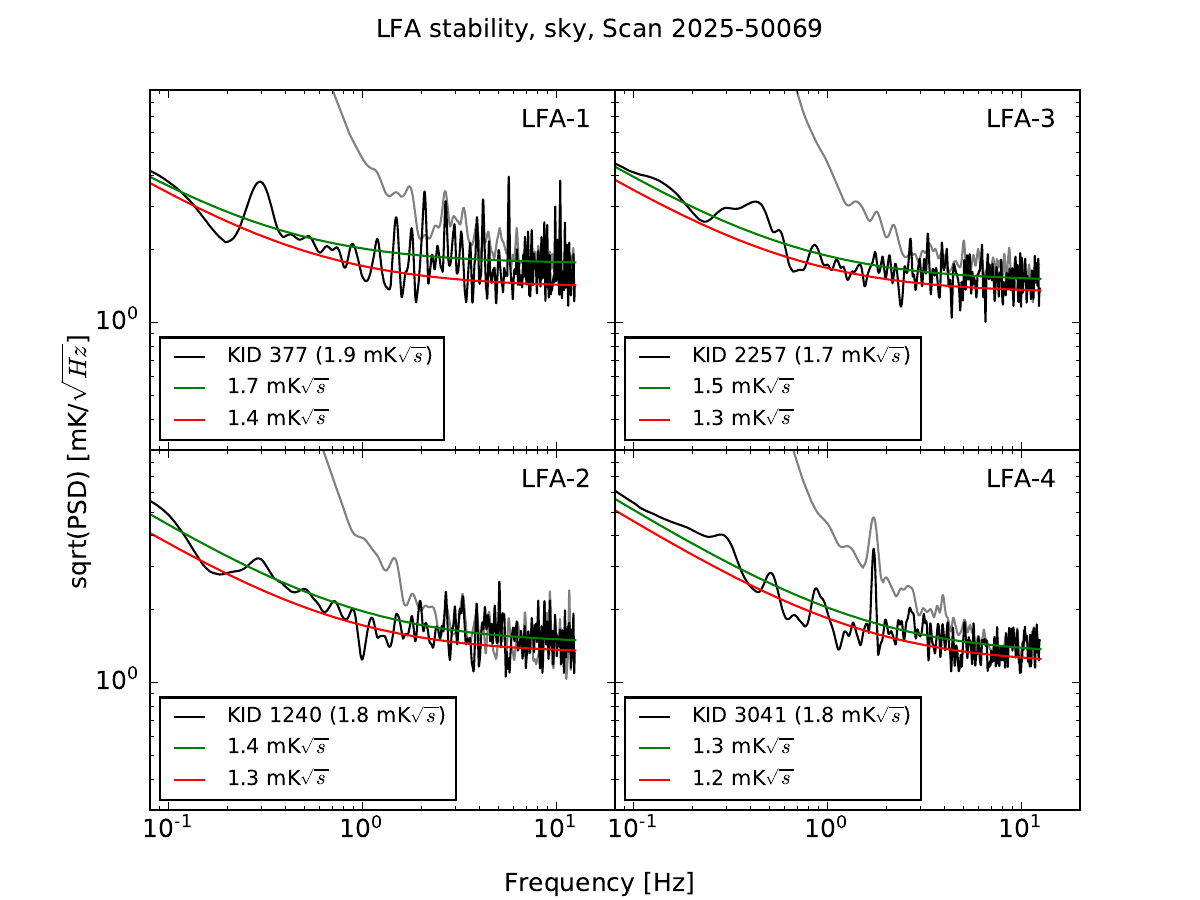}
      \caption{Noise spectra of typical LFA pixels in different readout chains. The data are from a 5-minute integration on a sky position. Gray traces show raw data; black traces show data after atmospheric noise removal. Both traces are smoothed with a 0.1~Hz Gaussian. The green line shows the best-fit to signal, indicating noise level above 2~Hz. The red line shows the estimated detector noise after median filtering of contaminant before data smoothing.}
         \label{fig:Stability}
\end{figure}

System stability is studied by analyzing the power spectra from a 300~s noise scan on blank sky with a static telescope. The noise trace is decorrelated to remove atmospheric fluctuations, then analyzed. Fig. \ref{fig:Stability} presents the noise power spectrum of four representative detectors, one per sub-array, with NET near the peak of the distribution. To facilitate visualization the power spectra are smoothed to 0.1~Hz. The figure shows the traces before and after atmospheric noise removal. The results indicates a 1/f knee point of 0.5~Hz, identical to that from the independent pre-integration chip-level measurements discussed in Sect. \ref{ArrayCharacterization}.\\
\indent The full resolution spectra allow us to disentangle the different noise sources affecting detector sensitivity. Due to 1/f contribution, the measured NET exceeds the plateau noise level above 2~Hz. This last value, indicated by the green line, is expected to be comparable with the photon noise level predicted by our simulations. This 1/f contribution arises from two sources:the inherent detector 1/f noise and the residuals from sky fluctuation removal.

The second noise source is unwanted contaminating signals, mainly at 0.3~Hz and its harmonics up to the maximum frequency. To estimate its relevance, we removed them from the spectra by using a median filter on the PSD before the smoothing process. The result, shown as the red line in Fig.~\ref{fig:Stability}, represents the estimated photon noise plus detector noise for each sub-array.

The above analysis indicates a contribution of around 10\% to the NET budget from contaminating signals in the spectra and 15\% contribution from 1/f noise. Additionally, we account for 15\% degradation from readout contribution. The values discussed here have a wide spread as many factors add up. As an example, high-band readout detectors have larger noise contributions from digitization than low-band detectors. Likewise, the 0.3~Hz signal is not uniform across the array, being much stronger toward the edges of the detector chips. With these considerations, we conclude that the photon noise level of our detectors accounts for almost half the measured noise.

This exercise is fundamental for instrument development, as it shows pathways for further improvements in coming years. Of special relevance is deploying a new-generation readout, which promises to lower digitization noise contribution and diminish the observed NET spread. Another improvement is suppressing the 0.3~Hz noise. This noise signal was initially observed during laboratory testing of the instrument, see Appendix \ref{Ap:Magnetic}, as a strong feature in the time lines. Adding magnetic shielding to the instrument attenuated this signal around 40~times, but as discussed above, it persists. Although the exact source of the signal remains unknown, the signal could be further attenuated by improving the instrument's magnetic shield.\\
\subsection{Optics performance} 
\begin{figure}[ht]
   \centering
   \includegraphics[width=\hsize]{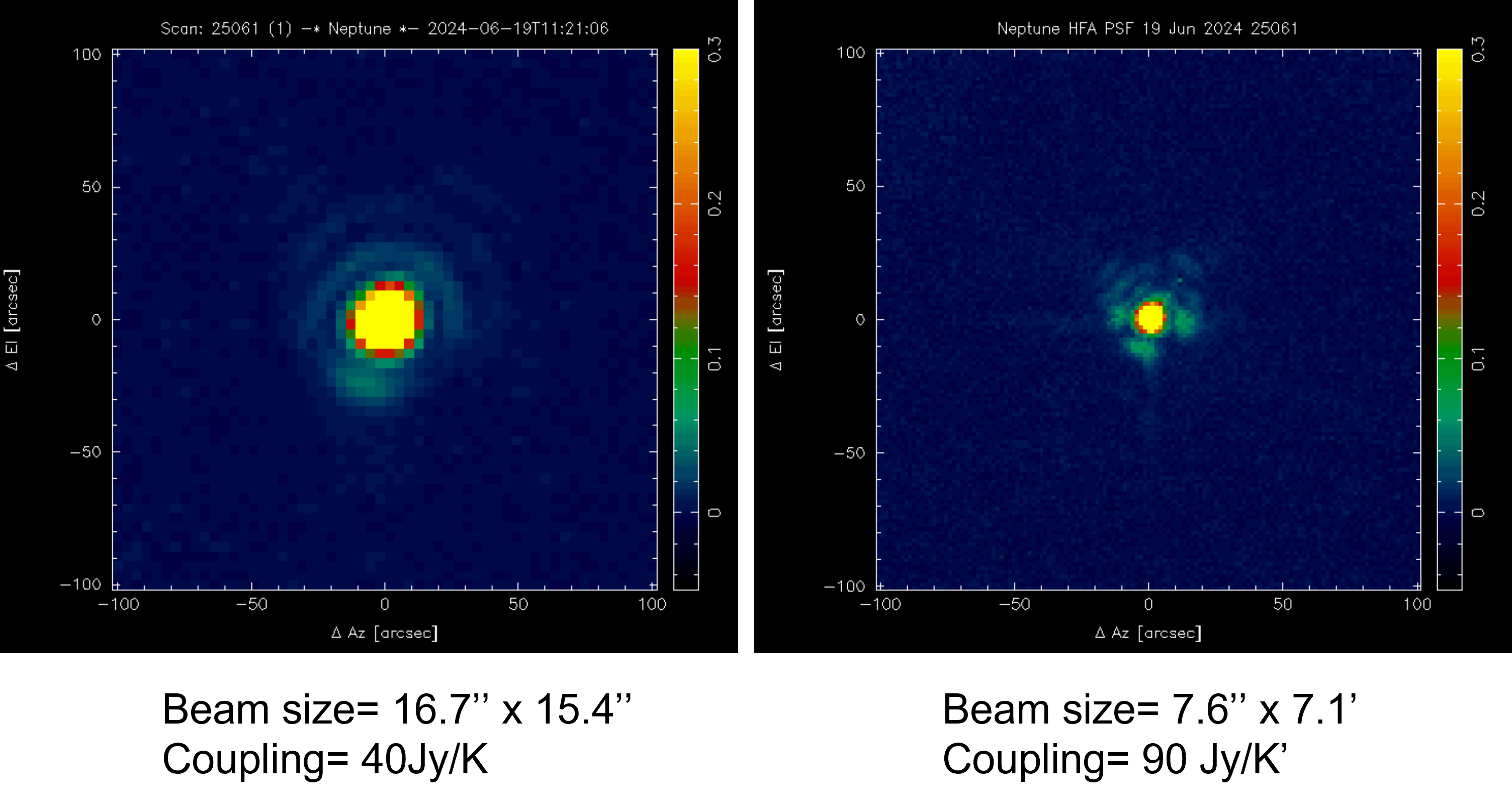}
   \caption{Beam measurement on Uranus (left: LFA; right: HFA). To make the error beam clearly visible, the color scale is saturated at 30$\%$. }
   \label{fig16}
\end{figure}

\begin{figure}[ht]
   \centering
   \includegraphics[width=1.0\hsize,trim={0cm 0cm 2cm 0cm},clip]{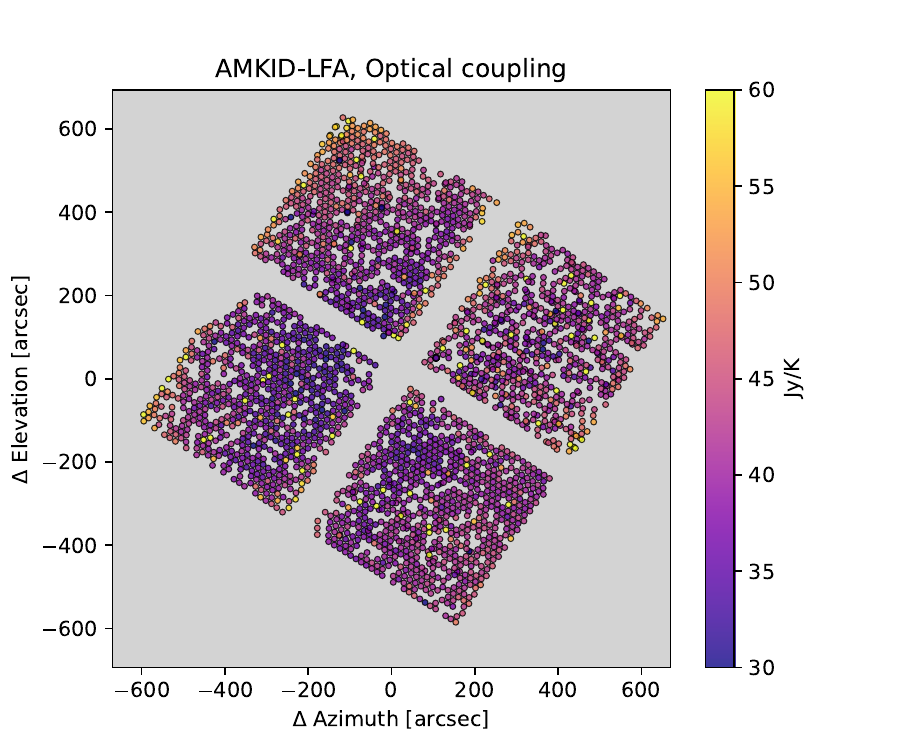}
      \caption{Beam measurement on Uranus. The image shows the measured optical coupling of AMKID-LFA. The average value is 40 Jy/K, corresponding to 61\% aperture efficiency.}
         \label{OpticalCoupling}
   \end{figure}

\begin{figure}[ht]
   \centering
   \includegraphics[width=\hsize]{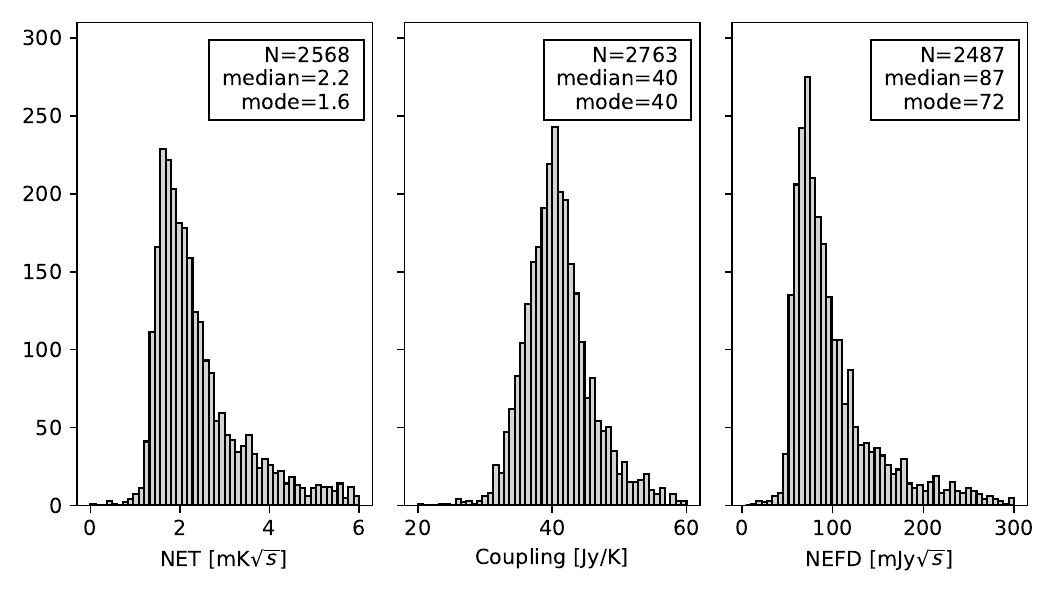}
      \caption{Typical distribution of measured NET, optical coupling, and NEFD for the 2025 observation season. }
         \label{fig:Hystograms}
   \end{figure}

Early results show that the instrument optics perform according to expectations. Planet scans, as those in Fig. \ref{fig16}, are regularly used to measure beam properties. Average beam sizes of 17.5'' with beam ellipticities lower than 1.1 are regularly measured with the LFA.\\
\indent A relevant optical parameter is the detector coupling to a point source ($C_{optic}$), i.e., the conversion factor from temperature to source flux in Jansky. It is defined as
   \begin{equation}
    \label{eq:OpticalCoupling}
      C_{optic}= \frac{10^{26} * 2 *  k_{b} } { A_{t}  \eta_{a} } \quad [Jy/K],
   \end{equation}
where $A_t$ is the telescope collecting area, k$_{b}$ is the Boltzmann constant, and $\eta_{a}$ is the aperture efficiency of the instrument and telescope, including Ruze efficiency. The factor of 2 accounts for total flux in double-polarization mode. Optical coupling is simply measured by comparing the expected flux from a calibrator -- usually a small planet such as Mars in opposition or Uranus -- to the measured KID response on its temperature scale. As expressed via Eq. (\ref{eq:OpticalCoupling}), it corresponds to a direct measurement of instrument aperture efficiency. Measurements in Fig. \ref{OpticalCoupling} and \ref{fig:Hystograms} indicate an average optical coupling of $40$~Jy/K at 870~$\mu m$, corresponding to 61\% aperture efficiency. We expect current efforts to improve the optical quality of the system, characterized by a half-wavefront error of 26~$\mu$m RMS \citep[see][]{Reyes_2023}, to further improve this value to around $37$~Jy/K.

In Fig. \ref{OpticalCoupling} the distribution of optical coupling along the field of view is color-coded. Measured beam sizes are shown as ellipses. The LFA presents a uniform response over the array with some edge degradation, especially on chip LFA-2, which is currently under study. Performance was initially measured after instrument installation and has remained stable for over a calendar year. \\
\subsection{Detector yield}
The first step in instrument operation is performing a KID search to identify usable resonances for observations. The list of resonances is stored by the backend for use in subsequent observations. The first definition of detector yield is thus the number of existing detectors correctly identified. The main limitation is that resonances are not equispaced in frequency and tend to cluster, as shown in Fig.~\ref{PowerSweep}. When two resonances are located at a distance lower than a resonance width, they begin to interact and exhibit a coupled response, as shown in \cite{Bisigello2016SPIE}. In extreme cases, resonances are so close together that they cannot be distinguished given our readout frequency resolution. Our algorithm uses a threshold of 0.15~MHz, which places a tone in only one of the two resonances when a collision is detected. Also, cross-talking pixels that survive this initial filter produce deteriorated NET and beam shapes and are thus filtered out in subsequent steps. 

After correct identification of resonances, the wire scanner is used to get an initial estimate of beam positions on sky and NET. Detectors with sensitivities larger than a certain threshold are not useful and are thus removed. This causes a second reduction in yield. Note that, as wire measurements are insensitive to beam shape, we reject pixels with poor chip-level performance. The main offenders are degraded lens array performance toward the edges of some sub-arrays, cross-talking detectors not removed in the previous step, and increased readout contribution at high frequencies.

Accurate detector positions on the sky are determined from on-the-fly (OTF) maps of Mars or Uranus, with parameters chosen to fully sample the planet for each detector (beam maps). These maps provide relative positions, optical coupling (measured by comparing the expected planet flux to the measured temperature calibrated via wire scanner), and beam ellipticity for each detector.

The noise equivalent flux density (NEFD) for each detector can be estimated by simply multiplying the NET by the measured optical coupling. The results allow us to derive the final instrument yield, defined as detectors with sensitivity in the target range. The results presented in Table \ref{table:Yield} show that the final on-sky yield of the LFA is around 71\%. \\
\begin{table}[ht]
\caption{AMKID-LFA detector yield.}
\centering
                
\centering                        
\begin{tabular}{l  c}     
\hline\hline               
Array & LFA \\
\hline                
   Number of pixels          & 3520   \\
   Detected resonances       & 3030 (86\%) \\
         NET in range    & 2568 (73.0\%)  \\
         NEFD in range   & 2487 (71.0\%)  \\
\hline \hline                           
\end{tabular}
\label{table:Yield}    
\end{table}
\section{Instrument early results}      \label{On-Sky}
Several verification and early science observations were carried out during the 2024 and 2025 observing seasons. The campaign verified that under good atmospheric conditions, the calibration routine is stable and enables observations for several hours without repeating the frequency sweep calibration. We consider good atmospheric conditions a PWV lower than 1.0~mm for the LFA. The need for recalibration is assessed via regular wire scanners and observations of calibration sources to monitor instrument sensitivity. Our observing strategies rely heavily on those developed for LABOCA (\citealt{Siringo2009}). The strategies include mapping routines to determine telescope foci as well as spiral and raster scanning patterns. Scanning speed is limited by the antenna control unit's position sampling rate of 20.8~Hz (below the AMKID readout sample rate of 30~Hz) to roughly 2.8 and 1.3~arcmin/s for the LFA and HFA, respectively.

We corroborate that determining the focus on the central area of the array is representative and that no degradation occurs due to focus settings toward the edges of the camera. A telescope pointing campaign using the HFA, demonstrates that the pointing model is stable and requires no major correction due to elevation tilting of our massive optic system. \\
   \begin{figure}[ht]
   \centering 
   \includegraphics[width=\hsize]{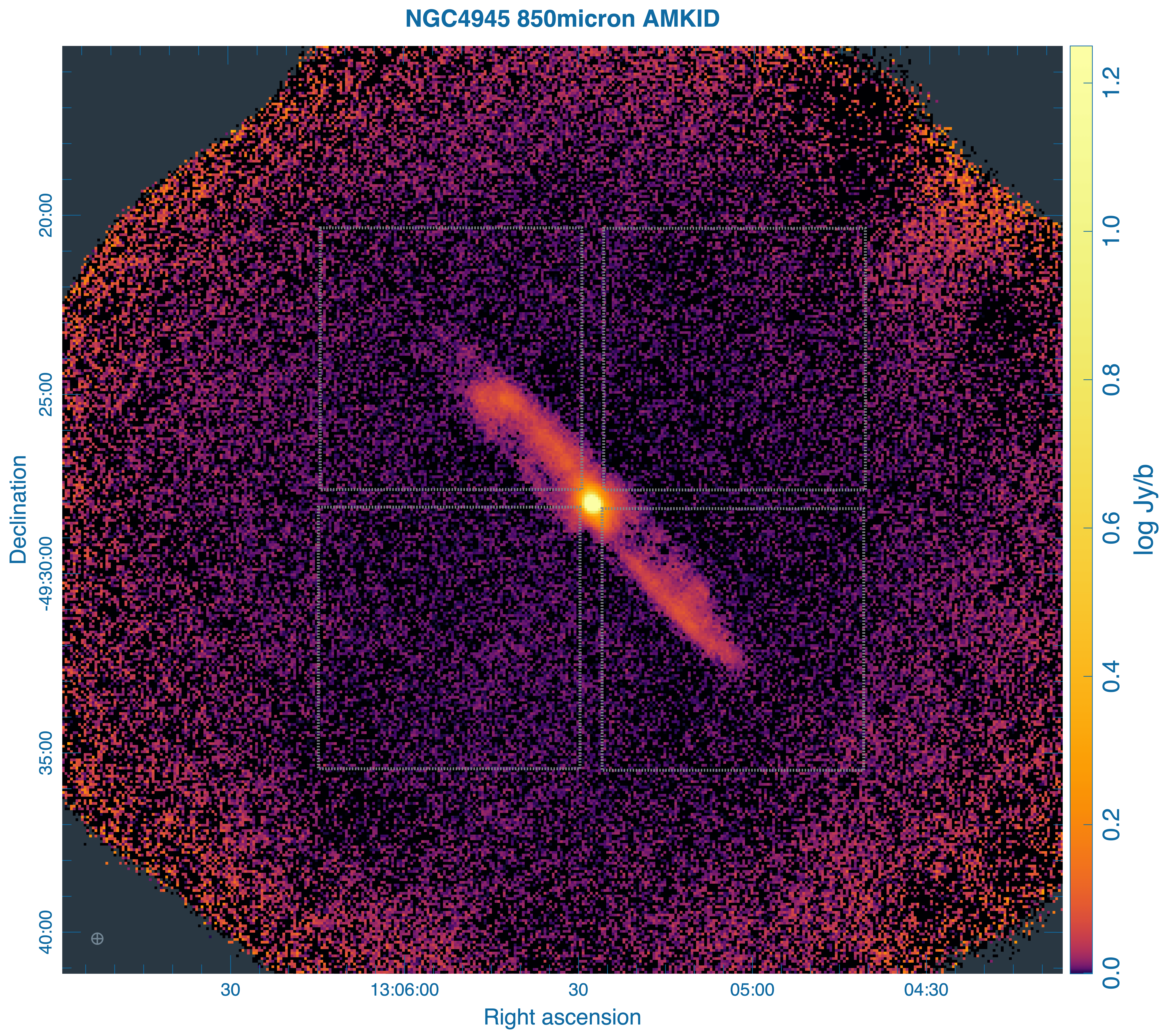}
         \includegraphics[width=\hsize]{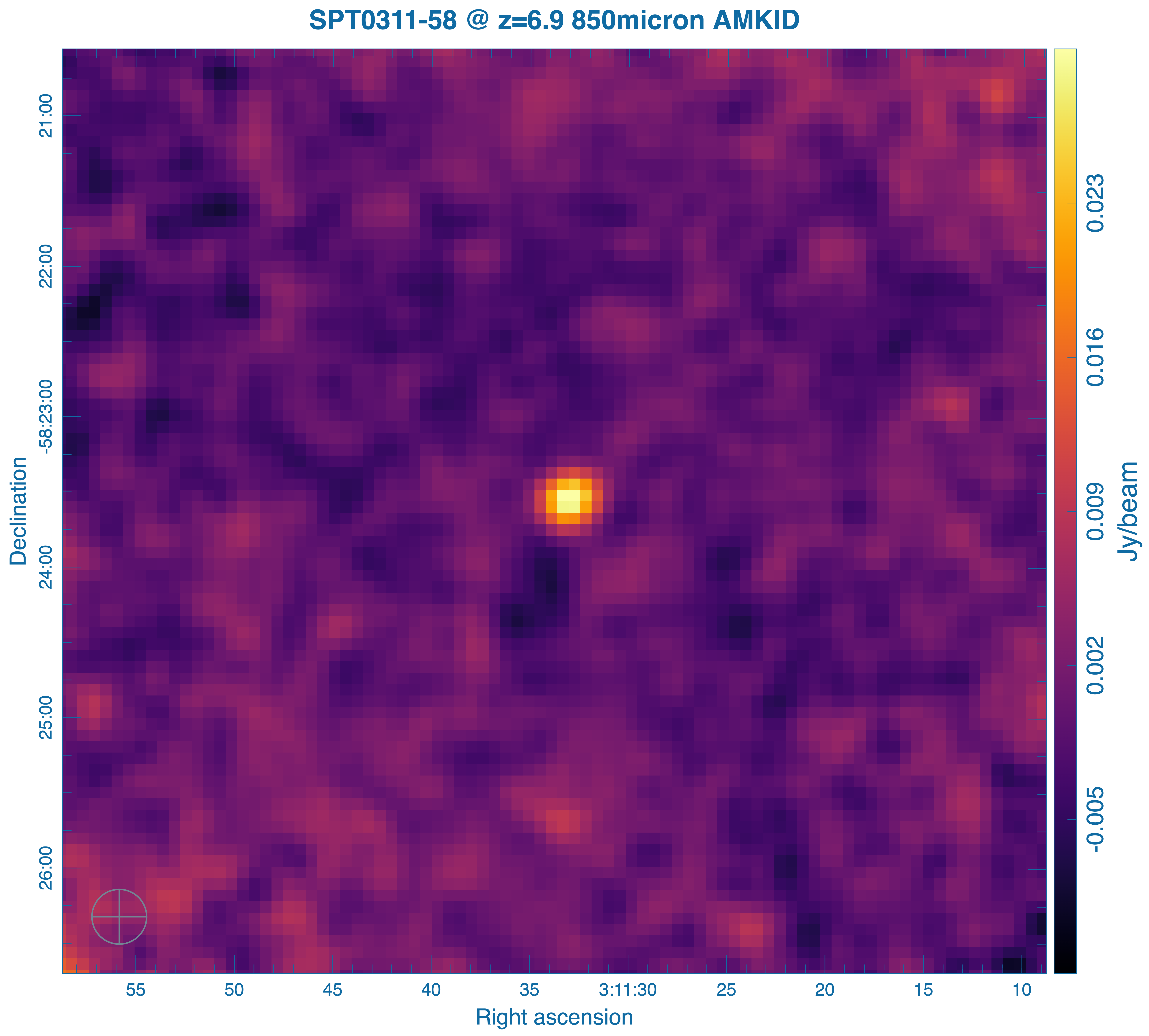}
      \caption{Top: AMKID-LFA image of the starburst galaxy NGC4945 at a distance of 3.8Mpc. The noise level of the image is 3.5mJy. Bottom: Zoom-in on the AMKID-LFA image of the dust star-forming galaxy SPT0311-58 at redshift 6.9, \cite{Strandet_2017}. The noise level of the image is 2.1mJy. Both images are smoothed to a beam size of 20''. The footprint of the LFA chips is shown as by the dotted boxes in the upper panel.  }
         \label{NGC4945}
   \end{figure}
\indent After these initial verifications were performed with both arrays, we focused our efforts on LFA commissioning. This array is currently validated and ready for scientific operation. Flux calibration -- based on wire scanner gain measurements for individual pixels plus optical coupling from planet scans -- proves a successful strategy, yielding typical calibration errors below 10\%. Sensitivity verification using calibration sources shows that on-sky sensitivities are 70-90~mJy$\sqrt{s}$ per beam, matching predictions from NET and optical coupling measurements. We expect large-scale structure filtering on scales corresponding to roughly half the chip size (3.5~arcmin), though details will depend on reduction steps, sky stability, and source structure. 

When mapping a square area of size $X_{scan}$ and $Y_{scan}$ with an integration time of $t_{int}$, the final RMS noise ($\sigma$) can be calculated using
   \begin{equation}
    \label{eq:IntegrationTime}
      \sigma= \frac{f_{samp}NEFD_{eff}}{T_{atm}}* \sqrt{\frac{(X_{scan}+\Delta)(Y_{scan}+\Delta)}{t_{int}A_{beam}N_{eff}}},
   \end{equation}
where $NEFD_{eff}$ is the median of the measured NEFD, ${N_{eff}}$ is the effective pixel number of the instrument, and T$_{atm}$ is the atmospheric transmission. The array size $\Delta=15.3'$ accounts for extending the mapping area to ensure homogeneous noise coverage at the edges of the map. The sampling factor ($f_{sam}$) is usually set to 2 to obtain Nyquist-sampled maps. Finally, the beam area ($A_{beam}$) can be easily calculated from the instrument's nominal beam size (17''). Using this formula the instrument mapping speed can be calculated as 25mJy (including 25\% overhead) per square degree per hour of integration under nominal weather conditions of 1.0~mm PWV. 

As an example, we show in Fig. \ref{NGC4945} an image of the nearby (3.8\,Mpc) starburst galaxy NGC4945 obtained with 2-hour on-source integration time \citep[compare with][]{Weiss_2008}, as well as the high-redshift dusty star-forming galaxy SPT0311-58 at z=6.9 \citep{Strandet_2017} with an on-source integration time of 2~hours.  Results indicate that the AMKID-LFA is 3.8~times faster than LABOCA at APEX when mapping extended sources. When comparing both instruments, it is important to consider their different architectures: while LABOCA was a horn-coupled array (2.5F$\lambda$) in dual polarization, AMKID is an over-sampled array (1.2F$\lambda$) in single polarization. This explains the difference in detector sensitivities: 55~mJy$\sqrt{s}$ for LABOCA detectors compared with 70-90~mJy$\sqrt{s}$ for AMKID detectors. The use of smaller and, therefore, noisier detectors, is compensated by the high number of detectors covering the field of view, as discussed in \cite{Griffin_2002}. Considering point-source sensitivity, number of detectors, and field of view (LABOCA had 290~detectors in a 12'' circular field), Eq. \ref{eq:IntegrationTime} predicts AMKID is 4.4~times faster than LABOCA. 

AMKID data can be processed using the {\it MARS} software package. {\it MARS} builds on the Bolometer array data Analysis (BOA) software package developed for LABOCA data reduction, but has been redesigned to support parallel data processing on multi-core machines and include all AMKID-specific reduction and calibration steps described here.\\
\section{Conclusions} 
The revised AMKID camera was successfully installed at the APEX telescope. Performing with an unprecedented number of pixels,  dual-color observing capabilities, high sensitivity, large field of view, and high angular resolution, this instrument is expected to open a completely new range of scientific opportunities with APEX in coming years. In this paper we presented an overview of the instrument design, with a special focus on science commissioning of its low-frequency channel. Early results indicate that the camera achieves typical sensitivities of around 70-90~mJy$\sqrt{s}$ per beam at 350~GHz. 

Dual-color operation of AMKID offers an additional high-frequency channel at 850~GHz. Initial observations using the AMKID-HFA demonstrate its high-frequency channel capabilities but also highlight the need for higher-quality optical components for successful scientific operation. The new optics system, consisting of a new set of warm mirrors with lower RMS error than the current system, will be installed soon. With this improvement early HFA science results are expected by the end of the 2026 submillimeter observing season. Additionally, a new readout currently under development promises overall improvement in sensitivity through lower readout noise and finer tone placement, thereby improving current detector yield.\\
\begin{acknowledgements}
The Authors thank T.M. Klapwijk for his work on the early development of the Aluminium-NbTiN hybrid KIDs.\\
This work is dedicated to the memory of Karl M. Menten (1957--2024) whose enthusiastic support of the APEX project made the development of AMKID possible.\\
\end{acknowledgements}
\bibliography{references_short}
\bibliographystyle{aa}
\begin{appendix}
\section{Laboratory instrument characterization} \label{APENDIX1}
Before deployment at the telescope, the instrument was extensively tested in laboratory environment. Performed measurements and analysis include optical quality and alignment, instrument sensitivity, and stability, among others. A dedicated experimental setup was built to mimic the instrument operational conditions. The system includes a mechanical structure to tilt the instrument, equivalent to 60-90~deg of elevation at telescope; a wire scanner to generate small signals over a cold background load; and a broad band photonic continuous wave signal generator mounted on a X-Y stage allowing for beam characterization of the system, as described in \cite{Reyes_2023}. \\
Of particular interest is the wire scanner device, presented in Sect. \ref{DataProcessing}. It consist of two small wires that moves against the sky background. Position of the wires is accurately record using Precision Time Protocol (PTP) over Ethernet and synchronized with data acquisition. The wire size (0.5mm) and material(standard electronic wire, plastic coated) were chosen to generate a small 10~K signal against the cold background.

The wire scanner was originally developed for fast characterization of the instrument performance in laboratory environment. For this a cold load is placed behind the wires mimicking the sky background. The load was implemented as a large copper bath covering the complete focal plane (80x80~cm$^2$). The bucket lower side is externally coated with Stycast conductive epoxy and silicon carbide grains to act as a cold load. The bath is thermally insulated by placing it inside a larger 4~cm wall thickness Styrofoam box. The effective temperature of this cold load was measured using a KID detector together with a rotating grid. The grid switches between two standard temperatures, ambient load at 300~K and liquid nitrogen load at 80~K. Measurements indicate an effective temperature of the load of 120~K for the low frequency band 160~K for the high frequency band, elevated from 77K due to the Styrofoam losses. However, these values are similar to the expected sky temperature at the deployment site. 

The wire-scanner characterization method proves to be effective and very fast. It's main purpose is to measure the temperature scale conversion factor, and give a first estimation of detector sensitivities.  Even when not perfect, several optical parameters as beam sizes and ellipticity, can be also estimated from the wire response shape. The complete array can be characterized in a couple of minutes; as a consequence, this method was extensively used as a basic characterization routine during the instrument optimization phase. Several optical issues were detected and corrected using this method. \\
\section{Magnetic shielding} \label{Ap:Magnetic}
\begin{figure}[!ht]
   \centering
         \includegraphics[width=0.77\hsize]{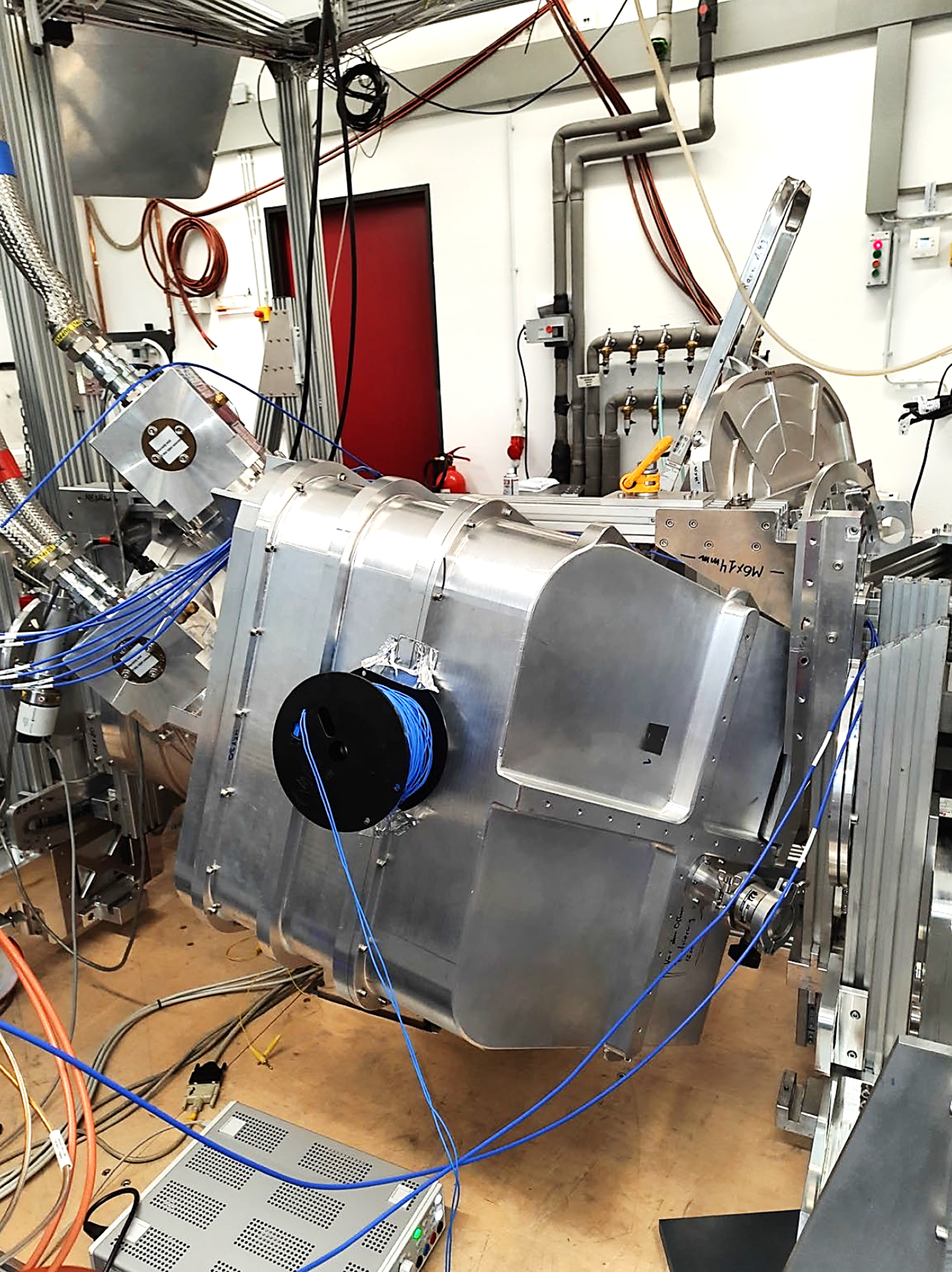}
      \caption{Photography of AMKID during laboratory testing phase. The coil seen on the instrument was used to generate external magnetic fields and study their impact in instrument operations. }
         \label{fig:Magnetic}
   \end{figure}

\begin{figure}[ht]
   \centering
         \includegraphics[width=1.0\hsize]{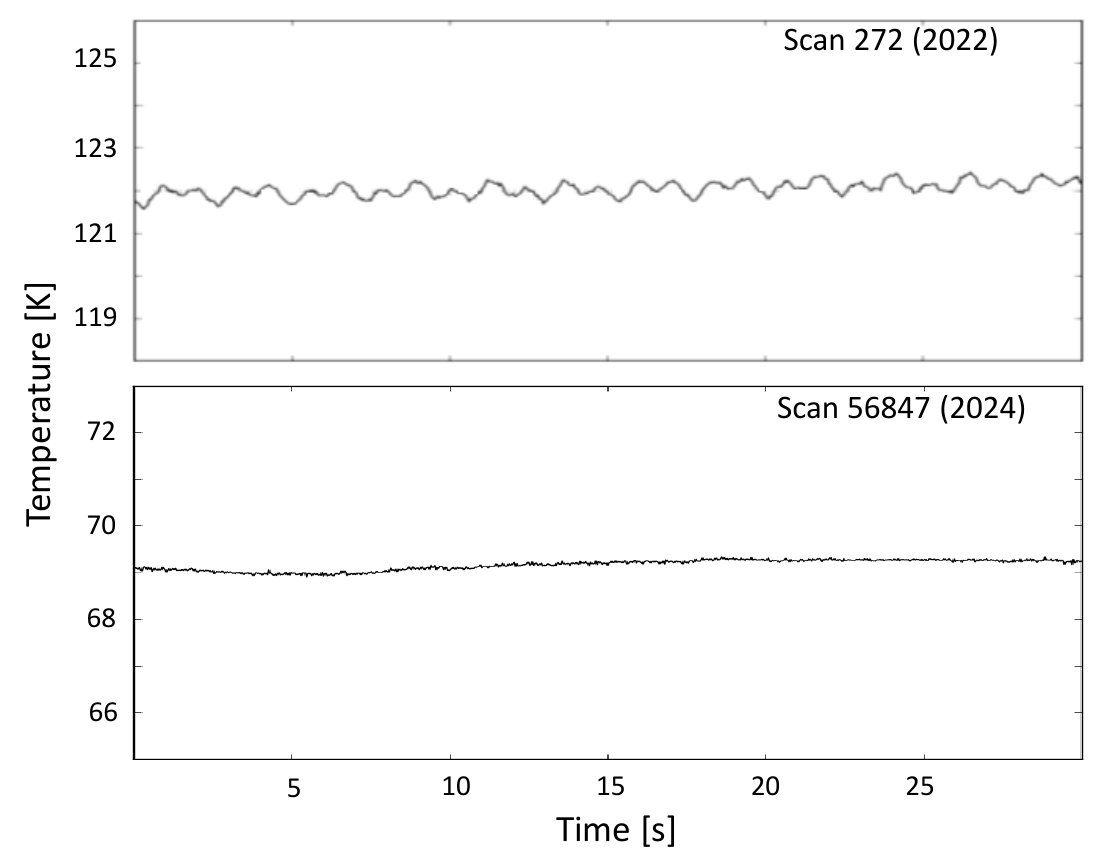}
      \caption{Top: Typical LFA timeline before the installation of the magnetic shield. Bottom: Timeline taken during instrument commissioning in 2024, i.e., after magnetic shield installation. The strong 0.3 Hz (and harmonics) signal present in the old data is now completely suppressed. The different in temperature level is due to old data being take in laboratory environment using the calibration load; meanwhile, the bottom panel data uses cold sky as background.  }
         \label{fig:Signal}
   \end{figure}

During instrument optimization it was clear that AMKID was extremely sensitive to external magnetic fields. This was specially prominent on the LFA, possibly due to the larger pixel size, almost 10 times larger than HFA pixels, enhancing the coupling of the detector to environmental magnetic flux. This phenomena was seen on various aspects of instrument operations:
\begin{itemize}
        \item Trapped flux during cool-down makes the resonances frequency to significantly move from one cool-down iteration to the next one.
        \item The low frequency array presented a prominent 0.3~Hz signal over most of the array, but specially strong at chip edges. The amplitud of the signal was as large as 0.5~K, as seen in Fig \ref{fig:Signal}. This signal is most likely magnetically coupled to the detectors.
        \item Initial testing of the instrument at the telescope in 2016 showed a strong signal of $\Phi=0.2$~rad (around 5~K) correlated with telescope azimuth movements. This was caused by the varying coupling to the earth magnetic field.
        \item Visible KID response is obtained by waving a ferromagnetic material (scissor, screwdrivers) around the cryostat.
\end{itemize}
This behavior presented a challenge for operations, but as induced signals are slow and correlated over the array they can be partially filtered out in post-processing. A more serious problem is the hysteric response of some detectors. This is seen as a sudden jump of level of the timelines of some detectors, usually happening when magnetic flux exceeds a threshold as described in \cite{Stan_2004}. This high frequency content signal cannot be filtered out and the affected detectors cannot be used for astronomical observations. This topic was investigated experimentally in laboratory environment by mounting an external small coil on the cryostat ( see Fig.~\ref{fig:Magnetic}) to artificially generate magnetic field as expected in regular operations. The main conclusion of the study is that the HFA performs well when facing magnetic fields up to 15~$\mu$T (maximal change during an Azimuth scan due to earth magnetic field). For the same field intensity the LFA presents a hysteric behavior. This noticeable difference is due to the increased pixel size at low frequency, giving a larger focusing of flux into the sensitive part of the KIDs. Other possible reasons for this difference are  the higher kinetic inductance of the LFA KIDs due to the thinner aluminum film, and the slightly worst shielding of the LFA by the aluminum cup.\\
\indent The AMKID original design included some magnetic shield consisting of an aluminum superconductive cup, with a moderate aspect ratio of four, located around each of the two arrays. Additionally, the detector chips include a superconductive back-plane which also act as a shield for detectors located at center of chip, but deteriorating edge detectors. The laboratory experiment shows that this shield was not sufficient and that a factor 40x of additional magnetic shielding was required to avoid jumps in timelines. This was addressed by including  a passive magnetic shield around the cryostat outside. It consists of a series of 1~mm thickness plates made of high magnetic permittivity metal ($\mu$ around 50000$\mu_0$), covering a large fraction of the cryostat. The coverage was carefully designed using numeric simulations for easiness of installation, meanwhile achieving the required attenuation at all possible orientations of the cryostat. After shield installation the above mentioned affects are no longer visible and a typical external field attenuation of 100 is observed during operations. \\
\section{The readout noise contribution} \label{DigitalNoise}
\begin{figure*}[ht]
   \centering
   \includegraphics[width=1\textwidth]{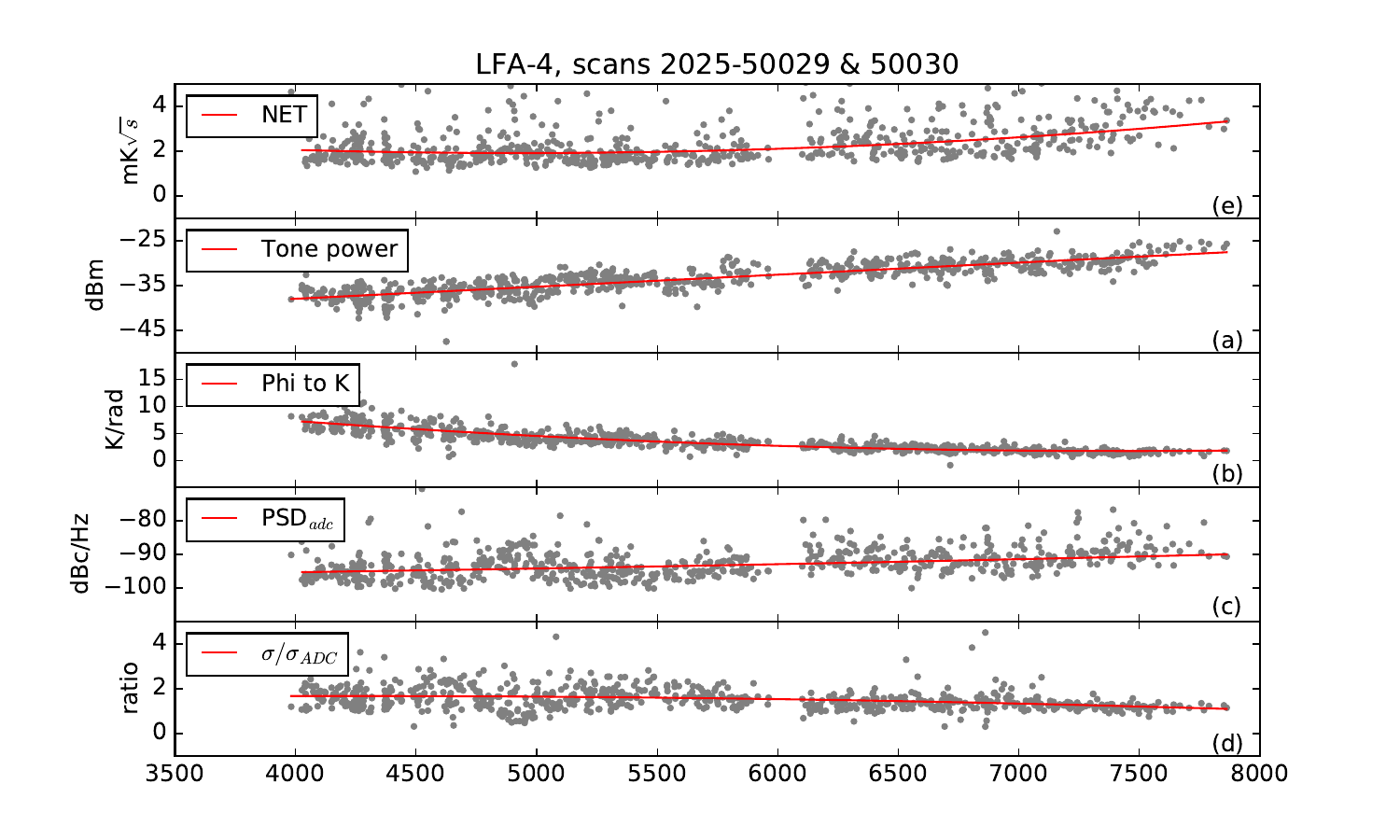}
      \caption{Frequency distribution of some discussed parameters of the KID readout system. }
         \label{fig:PSD2}
   \end{figure*}
The digitization noise caused by the finite number of bits in the ADC and DAC can be expressed by the PSD of the read tone normalized by the off-resonance tone power as
\begin{equation}
    PSD_{ADC}=N+C-6\mathrm{ENOB}-1.8+3-10 \log{f_s} \quad [dBc / Hz ]
                \label{eqn:PSD}
,\end{equation}
where $f_s=2.2$~GSamples/s is the sampling rate of the ADC; $N=10\log_{10} (n)$ is the noise degradation due to the number of tones (n) used by the readout; $C$ is the crest factor, the ratio between the peak voltage and the root mean square voltage of the waveform. We use a random initialization of tone phases achieving a typical crest factor of 14~dB. The factor 3~dB is because of the the double sideband down converter, which increases the noise floor by a factor of 2. The effective number of bits (ENOB) is a measure of the real performance of the digitization hardware and is lower than the real number of bits. The AMKID readout uses a 14-bit DAC and a 10-bit ADC, being the total PSD of the system dominated by the ADC noise, characterized by an ENOB of 7.6dB.

For 1000 tones the readout achieves a $PSD_{ADC}$ of -93.8~dBc/Hz, assuming flat power across the band. 

The system noise analysis shows that our readout noise is dominated by the above described digitization noise and not by the analog section of the readout, being the last more than 2 order of magnitude smaller that the digitization noise (-125.2~dBc/Hz).

\begin{figure}[ht]
   \centering
         \includegraphics[width=0.8\hsize]{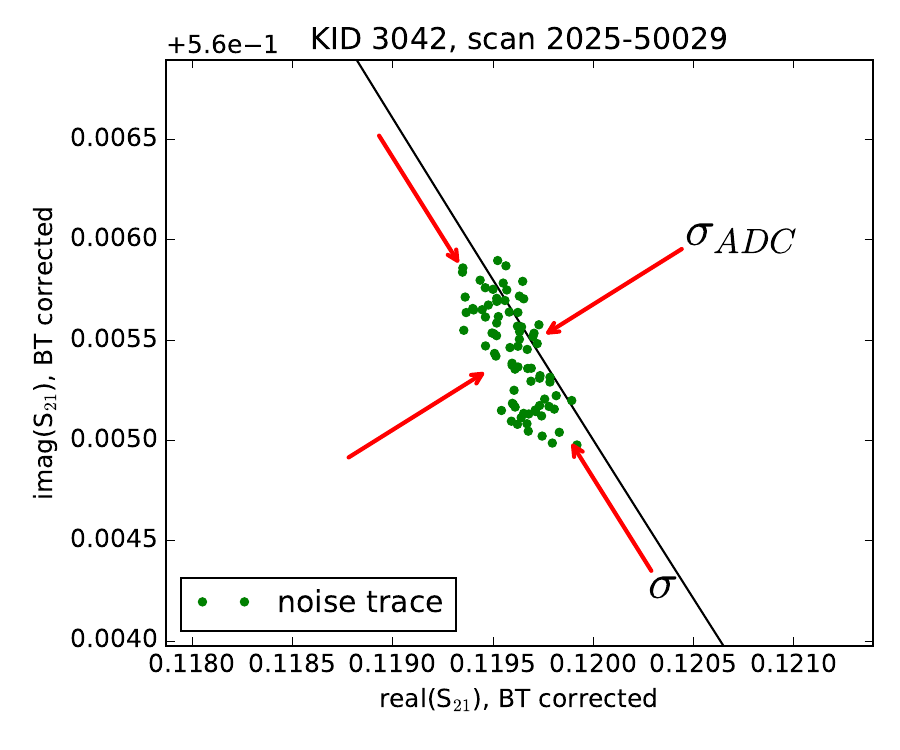}
      \caption{Five-second noise trace shown as IQ blind-tone corrected data in the S$_{21}$ plane. The solid black line show the KID circle. The spread along the circle tangent direction ($\sigma$) is caused by all source of noise in the system. Noise in the radial direction ($\sigma_{ADC}$) comes mostly from the readout. }
         \label{fig:PSD}
   \end{figure}

The digitized signal is integrated in a circular buffer and later Fourier transform as explained in Sect. \ref{Readout}. The output signal from the digital section is the gain coefficient associate to each readout tone. We can express the digitization noise as a standard deviation of the measured gain coefficient in a one second measurement as
\begin{equation}
    \sigma_{ADC} = \frac{1}{\sqrt{ N_{samp}  N_{int}}} \sqrt{\Delta_{chan} 10^{PSD_{ADC}/10}}    \quad 
                \label{eqn:IQ_noise}
,\end{equation}
where  $\Delta_{chan}$ is the channel width, $N_{int}$ the number of integrations in the circular buffer, and $N_{samp}=f_{samp}$ the number of samples per second. The above expression can be further simplified as $\Delta_{chan}=N_{samp} N_{int}$. The measured standard deviation of the S$_{21}$ can be temperature calibrated following the procedure described in Sec. \ref{DataProcessing} and then expressed as a contribution to the system NET:
\begin{equation}
    NET_{ADC} =  \frac{\sigma_{ADC}}{R} K \frac{1}{\sqrt{f_{samp}}}  \quad [K \sqrt{s}]
                \label{eqn:readoutNET}
,\end{equation}
where K is the conversion factor from phase shift ($\phi$) to temperature scale as discussed in Sect. \ref{DataProcessing}; f$_{samp}$ the sampling frequency; and R is the radius of the KID circle, which is related to the KID depth (min(S$_{21})$) as 
\begin{equation}
    R =   \left( \frac{1-min(S_{21})}{2} \right) \quad 
                                \label{eqn:KIDCircleRadius}
.\end{equation}
During observations the noise performance of the system is regularly monitor by executing noise scans after a calibration sweep. This data is analyzed and the $\sigma_{ADC}$ measured as the standard deviation of the $S_{21}$ data along the KID circle radial direction as shown in Fig.~\ref{fig:PSD}. At the same time the standard deviation along the tangent direction is computed as $\sigma$. The measured NET is proportional to $\sigma$, meanwhile the digitization NET contribution is directly proportional to the $\sigma_{ADC}$ value. This fact is used to estimate the relevance of the NET contribution to the total measured NET using the following expressions:
\begin{equation}
    \frac{NET_{meas}}{NET_{ADC}} =  \frac{\sigma}{\sigma_{ADC}},
\end{equation}
\begin{equation}
                NET_{meas}=\sqrt{NET_{system}^2 + NET_{ADC}^2}.
\end{equation}
In our system we obtain a typical separation (ratio between $\sigma_{ADC}$ and $\sigma$) of 2.8~dB, meaning that the ADC contribution is on the same order of magnitude than the instrument NET. Giving the quadratic adding of NET this means that the ADC noise accounts for a 15\% of the measured NET, indicating a clear path to improve the system performance by using a better readout with a larger ENOB.
\section{NET modeling} \label{A:NET}
The amount of power per unit of frequency that reach the detector, $P(\nu)$, can be divided in two main sources: The power arriving from the sky background, $P_{sky}(\nu)$, and the power that is generated by the filter stack itself, $P_{filter}(\nu)$. As filter stack we consider all elements that are placed from the cryostat window to the last bandpass filter, including the polarization grid and the cold-stop. Note that the lens array is not considered as part of the filter stack. 
\begin{eqnarray}
P(\nu)=P_{sky}(\nu) + P_{filter}(\nu) 
.\end{eqnarray}
The $P_{filter}$ term is calculated using the contribution from each element in the filter stack. For this an iterative model is used: all elements are modeled by its physical temperature and its transmission, reflection and absorption coefficients. At each iteration the emitted, absorbed, reflected and transmitted power are computed for each filter and used to calculate the thermal balance of the system when exposed to the ambient temperature. The process is repeated until the system reach thermal equilibrium. As a result we obtain the physical temperature and emitted power of each filter. This information is used to finally obtain $P_{filter}$. 

On the other side, $P_{sky}$ corresponds to a blackbody spectra with an associated temperature $T_{sky}$ which is filtered by the filter stack. It is important to notice that both terms, $P_{sky}(\nu)$ and $P_{filter}(\nu)$, include in one way or another the polarization efficiency ($\eta_{pol}$=0.5), the cold stop efficiency ($\eta_{pupil}$),  and the window and filter physical sizes. They also include the transmission of all other elements in the stack, which for convenience we group in a term called $\eta_{filter}$. 

To obtain the power that reaches an individual detector we use the following expression:
\begin{eqnarray}
P_{KID} (\nu)=0.86 P(\nu)\frac{A_{detector}}{A_{array}} \eta_{lens}   
,\end{eqnarray}
which calculates the power that reach each detector as the ratio between the detector area ($A_{detector}=\pi R^2_{lens}$) to the full array area ($A_{array}=0.0144$~m$^2$) multiplied by the lens-antenna radiation efficiency ($\eta_{lens}$) and the coupling of the lens-antenna to a plane wave (86\%). This last factor is used as our lens-antenna system produces a gaussian electric field distribution at lens aperture and not an uniform distribution as expected for noise photons.

The number of photons per unit of frequency that reach a detector during a second is calculated as\begin{eqnarray}
n_{v}= \frac{P_{KID}(\nu)}{h\nu}  
.\end{eqnarray}
From here we calculate the Noise Equivalent power (NEP) of a single detector as \begin{eqnarray}
NEP= \sqrt{2\int {(h \nu)^2  n_{v}  (1+ n_{v} + \frac{2\Delta}{\nu \eta_{qp}})d\nu} }    \quad [W\sqrt{Hz}]
.\end{eqnarray}
Expression that include the Poisson  noise together with the wave and recombination noise contributions, see \cite{Flanigan_2016} and \cite{Ferrari_2018a}. $\eta_{qp}=0.57$ is the efficiency of the quasiparticle generation and $2\Delta=90GHz$ is the superconductor bandgap frequency. The factor $\sqrt{2}$ is used to express the NEP in units of $[W\sqrt{Hz}]$, i.e., as the S/N obtained in a 0.5~s integration time.

As last stage we convert the NEP measured at detector level to NET measured at instrument focal plane, value that is comparable with our measured NET using the wire scanner.
\begin{eqnarray}
NET_{instrument}= \frac{1}{\sqrt{2}}*\frac{1}{2 K_{b}  BW} * \frac{NEP_{detector}}{\eta_{filter}\eta_{lens}\eta_{pupil}\eta_{pol}}    [W\sqrt{s}]
,\end{eqnarray}
where $K_b$ is the Boltzmann constant and $BW$ the filter bandpass. The factor $\sqrt{2}$ is used here to express the NET in units of $K\sqrt{s}$, i.e., as S/N after a 1 second integration time. In our case the factor 2 in the denominator cancels with the $\eta_{pol}$, but it is explicit in the equation to facilitate extension of this procedure for dual polarization instruments.

All the values used in the above calculations had been discussed in the article main body. To facilitate the use of above equations we provide Table \ref{tab:NETParameters} showing a summary of the required values.
\begin{table}[ht]
\caption{AMKID NET calculation parameters.}     
\footnotesize  \centering
\begin{tabular}{c | c c }     
        \hline\hline               
        Parameter & LFA & HFA    \\
        \hline
                NET & 0.83 mK$\sqrt{s}$&1.16 mK$\sqrt{s}$ \\                                                                            
                $R_{lens}$ & 1.0 mm  &  0.55 mm  \\
                $\eta_{lens}$ &  0.84 & 0.76  \\
                $\eta_{pupil}$ &  0.45 & 0.55  \\
          $\eta_{pol}$ &  0.5 & 0.5  \\
          $\eta_{filter}$ &  0.56 & 0.41  \\
                $\nu_{0}$ & 345GHz & 850GHz \\
          $\Delta B$ &  38 GHz & 123 GHz  \\
        \hline\hline
        \end{tabular}
        \label{tab:NETParameters}
\end{table}

\end{appendix}
\end{document}